# Physics Letters B, Physical Review Letters and arXiv publications.

Correlating PLB, PRL and arXiv articles for nuclear, particle and astro-physics

*by*

*H. Weerts*
*Dept. of Physics & Astronomy, Michigan State University and*
*HEP Division, Argonne National Laboratory*

*July 2020*

**Abstract**

When analyzing the content of Physics Letter B (PLB) web pages, one goal was to separate articles from particle physics(HEP) and nuclear physics(NP). PLB contains information about the subject area of an article: Astrophysics & Cosmology, Experiments, Phenomenology or Theory. Those subject areas have been used since 2004. To extend those areas to earlier dates and try to separate HEP and NP publications, the idea was to use the information on the arXiv to accomplish this, by correlating a publication in PLB with a publication in one of the different arXiv repositories for *hep-ex, hep-ph, hep-th, hep-lat, astro-ph, nucl-ex* or *nucl-th.* The arXiv articles go back to at least 1995. When building the correlation between an arXiv article and each PLB article, it was found that many PLB articles do not have an existing reference in an arXiv article. Given that, the same analysis was performed using particle, nuclear and astro physics publications in Physical Review Letters (PRL). A similar result was obtained, but for a much shorter period of time. Finally, a title match was performed between the two journals and the arXiv, to see whether the "preprint" version of the journal article existed on the arXiv at all. They do, but without a journal reference.

**Contents**

## Introduction

The field of particle physics (HEP) has produced many articles over the last seven decades or more. These articles describe progress in theory, experiment and phenomenology. They have been published in an array of different journals. With the start of digital content about 30 years ago, all this information is easily accessible these days through web pages from journals. In addition the arXiv [1] , started in the early 1990's, has provided a platform to provide digital access to what was called "preprints" before that time. The arXiv contains early versions of articles that are published in refereed journals, and in addition to that a large compilation of articles (conference procedures, lectures, notes etc.) not necessarily published in refereed journals. Articles on the arXiv are assigned to research subject areas or repositories, which in principle distinguish HEP and NP and experimental, phenomenological and theoretical results in those areas (see more below).

The original purpose of this study, started last year, was to find all possible information about an article published in a refereed journal, including its history before it was published, which would include its history on the arXiv. Typically, a first version of an article is published on the arXiv. This article is then submitted for publication in a refereed journal. Once accepted for publication and published, the information on the arXiv is updated (as far as we know this is up to the author and not automatic), to include the reference to the published journal version. On the journal side the information, about from which arXiv article the journal publication originated, is lost and/or never published and not available on journal webpages. So given a refereed journal article it is not straightforward to find its source on the arXiv.

Journals have different ways of grouping their articles by field. For example, Physics Letters B (PLB) [2] introduced subject categories: *Astrophysics & Cosmology*, *Experiments*, *Phenomenology* and *Theory* in 2004. The subject areas contain articles spanning astrophysics, cosmology and particle physics (HEP) and nuclear physics(NP). Before 2004 there is no information about the subject area. By combining the subject areas in journals and knowing from which arXiv repository an article originates, it should be possible to identify whether a journal article is either astro, particle or nuclear physics and whether it describes experimental, phenomenology or theoretical results. The goal of this work is to establish a correlation between a published journal article and its preprint version on the arXiv, starting from the journal article and using the journal reference in the arXiv article. Once this information exists i.e. knowing from which arXiv repository a journal article originates one can separate PLB publications into HEP and NP.

Physical Review Letters (PRL) [3] serves the whole physics community and has many subject areas. To compare to the study described above, letter articles from three subject areas were selected from all PRL volumes over the years. They are *Gravitation and Astrophysics*, *Elementary Particles & Fields* and *Nuclear Physics*. For PRL for example one could then subdivide *Elementary Particles & Fields* articles into experiment, phenomenology and theory.

## Approach and data gathering

It should be clearly stated at the outset that the main goal of this report is to look at published journal articles in HEP and NP. This could be expanded in the future, but for now it is limited to those fields. This report only covers journal articles in PLB and PRL, but it will start with PLB, because historically that was the first journal looked at. We will return to PRL later.

A basic assumption is that any article submitted to a journal will first be submitted to the arXiv, in the relevant repository. After acceptance by the (refereed) journal, it is further assumed that the information in the arXiv article will be updated (presumably by the author(s)) and will now include the link to the journalpublished article. The arXiv repositories expected to contain articles submitted to PLB are Astrophysics (astro-ph), High energy Physics Experiment (hep-ex), High Energy Physics Lattice (hep-lat), High Energy Physics Phenomenology (hep-ph), High Energy Physics Theory (hep-th), Nuclear Experiment (nucl-ex) and Nuclear Theory (nucl-th). By finding the preprint version of a journal article on the arXiv it would be possible to "classify", for example, the PLB article as NP or HEP and further subdivide it into experiment, phenomenology or theory.

To accomplish this, scripts were written that downloaded and analyzed the monthly arXiv submission web pages for all the repositories mentioned above, from when the arXiv started until the end of 2019. This was done in January 2020 and that is the contents described here. In order to avoid articles appearing more than once, only arXiv publications were retained where the first subject area of the article is identical to the repository name. Articles on the arXiv can appear in multiple repositories, but the subject areas (equal to the arXiv repository) in each article are ordered and identical on all repositories. This way an article only enters once into our analysis.

We perform the following steps to get an inventory of all the arXiv articles:

- *Step 1*: Download all monthly webpage summaries for each arXiv repository. Small summary files are created with a line for each article and these are basically used for any further steps. This way the original arXiv webpage is only used once.
- *Step 2*: Select articles where the first subject area in the article is identical to the repository name.
- *Step 3*: Select those articles that have a link to an article in a published journal. This does not include "Submitted to xyz". Specifically look for the "Journal-ref" associated with the article. At this point the arXiv articles are split into a set with journal references (which will be referred to as *arXiv-with-journal*) and a set without journal references (referred to as *arXiv-without-journal*)

Table 1 contains the number of articles selected in the above steps for each of the arXiv repositories.

| ArXiv area | Total # manuscripts, after step 1 | After step 2 | After step 3 (*arXiv-with-journal*) | Number without journal entry (*arXiv-without-journal*) |
|---|---|---|---|---|
| *hep-ex* | 40507 | 19064 | 10429 | 8635 |
| *hep-ph* | 147339 | 109789 | 67669 | 42120 |
| *hep-th* | 134238 | 87832 | 57607 | 30225 |
| *nucl-ex* | 19855 | 9480 | 5653 | 3827 |
| *nucl-th* | 46060 | 28012 | 18504 | 9508 |
| *hep-lat* | 23068 | 15418 | 10470 | 4948 |
| *astro-ph* | 265024 | 234051 | 84140 | 149911 |

Table 1: Showing the resulting articles in each arXiv repository after the steps described above. The last column is simple the number of articles that remain after step 2 without a journal reference. It is basically "Step 2 minus Step 3".

The purpose for displaying the numbers in Table 2 is to give an indication of what is on the arXiv and what the input is to the remaining steps. The next step is to take the *ArXiv-with-journal* set for each repository and interpret the information stored in the journal reference. This is the reference to the journal where the arXiv manuscript is published. Interpreting that information is not quite straightforward, since authors use different ways to reference a publication. Different conventions are used and sometimes simple spelling mistakes throw off algorithms. There are at least two different conventions for the PLB reference and sometimes they change over time. An algorithm was written to identify most publishers referenced, including PLB and decompose the journal information into journal name, volume, page or article ID and year. Using the algorithm developed, it is possible to count the number of articles on the arXiv in each repository published, in a given year and in a certain journal. It should be noted that any such algorithm is never 100% perfect. We will come back to this later, but the algorithm developed here interprets more than 99% of the journal references in this arXiv data set correctly.

As an example Table 2 below shows this information for the repository *hep-ex* for all years available. Table 2 is just an illustration of the existing data for all repositories. This is a rather simple one. Other ones have more/different journals in which they publish. The journals are the standard journals and listed in Appendix A. Any journal not listed is contained in the column "Other". The year "9999" counts those articles where the year information was not found or could not be encoded. Tables like this exist for each arXiv repository used here. Appendix B shows the corresponding table for the repository *nucl-th*, just as an illustration on how the journals, where articles are published, change with repository.

These tables for all arXiv repositories analyzed, contain the information of how many articles on the ArXiv have been published in a journal, by selecting that column. Although many journals are identified, for now we will concentrate on the columns containing PRL and PLB. If all articles published in PLB for example are referenced on the arXiv, the number under PLB should correspond to the number of articles published in PLB in that year.

| Year | JINST | PRD | JHEP | EPJC | EPJA | PLB | PRL | PRC | New J Phys | NPB | NPA | China | JPhyG | Nuovo Cim | NIM | Astro | IJMP | Conf | Review | Acta | PAN | Other |
|---|---|---|---|---|---|---|---|---|---|---|---|---|---|---|---|---|---|---|---|---|---|---|
| 1992 | 0 | 0 | 0 | 0 | 0 | 0 | 0 | 0 | 0 | 0 | 0 | 0 | 0 | 0 | 0 | 0 | 0 | 2 | 0 | 0 | 0 | 1 |
| 1993 | 0 | 0 | 0 | 0 | 0 | 0 | 0 | 0 | 0 | 0 | 0 | 0 | 0 | 0 | 2 | 0 | 0 | 0 | 0 | 0 | 0 | 0 |
| 1994 | 0 | 2 | 0 | 0 | 0 | 7 | 6 | 0 | 0 | 0 | 0 | 0 | 0 | 0 | 2 | 0 | 0 | 1 | 1 | 0 | 0 | 3 |
| 1995 | 0 | 5 | 12 | 0 | 0 | 22 | 27 | 0 | 0 | 5 | 0 | 0 | 0 | 1 | 11 | 0 | 0 | 3 | 0 | 0 | 0 | 5 |
| 1996 | 0 | 12 | 14 | 0 | 0 | 18 | 18 | 0 | 0 | 6 | 0 | 0 | 2 | 1 | 14 | 0 | 0 | 15 | 2 | 0 | 0 | 4 |
| 1997 | 0 | 20 | 18 | 0 | 0 | 37 | 39 | 1 | 0 | 6 | 0 | 0 | 0 | 0 | 10 | 0 | 1 | 17 | 3 | 4 | 0 | 5 |
| 1998 | 0 | 32 | 0 | 43 | 1 | 54 | 48 | 3 | 0 | 3 | 4 | 0 | 1 | 0 | 18 | 1 | 4 | 27 | 3 | 0 | 1 | 6 |
| 1999 | 0 | 30 | 2 | 55 | 2 | 76 | 44 | 1 | 0 | 3 | 4 | 0 | 5 | 1 | 30 | 2 | 2 | 63 | 3 | 2 | 1 | 3 |
| 2000 | 0 | 57 | 0 | 67 | 1 | 118 | 64 | 2 | 1 | 2 | 11 | 0 | 5 | 0 | 42 | 1 | 27 | 45 | 5 | 3 | 5 | 10 |
| 2001 | 0 | 47 | 0 | 43 | 3 | 82 | 93 | 3 | 0 | 7 | 1 | 0 | 4 | 2 | 34 | 3 | 40 | 86 | 1 | 0 | 1 | 17 |
| 2002 | 0 | 59 | 0 | 27 | 0 | 100 | 77 | 4 | 0 | 3 | 8 | 0 | 12 | 0 | 42 | 6 | 16 | 109 | 1 | 14 | 4 | 16 |
| 2003 | 0 | 48 | 0 | 44 | 2 | 74 | 62 | 3 | 0 | 2 | 9 | 0 | 9 | 0 | 42 | 6 | 1 | 87 | 7 | 6 | 5 | 14 |
| 2004 | 0 | 92 | 1 | 87 | 1 | 81 | 100 | 5 | 0 | 4 | 0 | 0 | 5 | 0 | 20 | 5 | 11 | 48 | 1 | 9 | 4 | 21 |
| 2005 | 1 | 111 | 0 | 31 | 5 | 85 | 126 | 4 | 0 | 4 | 11 | 0 | 3 | 0 | 14 | 0 | 44 | 61 | 5 | 15 | 2 | 13 |
| 2006 | 3 | 132 | 1 | 44 | 1 | 65 | 109 | 3 | 1 | 1 | 3 | 0 | 2 | 0 | 8 | 0 | 19 | 113 | 5 | 16 | 4 | 14 |
| 2007 | 0 | 119 | 6 | 53 | 1 | 44 | 110 | 4 | 0 | 7 | 5 | 0 | 3 | 0 | 4 | 2 | 7 | 128 | 3 | 6 | 7 | 15 |
| 2008 | 1 | 132 | 10 | 26 | 1 | 46 | 123 | 6 | 0 | 4 | 0 | 0 | 5 | 13 | 3 | 4 | 5 | 103 | 8 | 11 | 2 | 16 |
| 2009 | 8 | 118 | 9 | 34 | 5 | 49 | 94 | 12 | 0 | 2 | 15 | 0 | 8 | 5 | 8 | 1 | 5 | 106 | 6 | 5 | 0 | 14 |
| 2010 | 2 | 102 | 15 | 28 | 1 | 44 | 68 | 7 | 0 | 4 | 2 | 0 | 1 | 2 | 8 | 2 | 3 | 98 | 5 | 3 | 3 | 13 |
| 2011 | 1 | 114 | 31 | 36 | 3 | 62 | 83 | 3 | 4 | 2 | 6 | 0 | 2 | 1 | 3 | 2 | 7 | 54 | 6 | 3 | 5 | 19 |
| 2012 | 2 | 96 | 72 | 41 | 0 | 94 | 84 | 4 | 0 | 1 | 0 | 0 | 0 | 0 | 2 | 2 | 2 | 19 | 10 | 4 | 0 | 13 |
| 2013 | 4 | 103 | 70 | 35 | 1 | 72 | 60 | 2 | 5 | 3 | 4 | 0 | 0 | 2 | 3 | 1 | 5 | 20 | 4 | 0 | 0 | 23 |
| 2014 | 2 | 120 | 72 | 39 | 1 | 43 | 71 | 5 | 2 | 2 | 4 | 0 | 0 | 0 | 4 | 0 | 5 | 20 | 10 | 0 | 0 | 20 |
| 2015 | 5 | 133 | 73 | 49 | 0 | 46 | 62 | 7 | 0 | 2 | 0 | 0 | 0 | 1 | 0 | 1 | 3 | 20 | 6 | 3 | 0 | 21 |
| 2016 | 5 | 125 | 72 | 66 | 1 | 70 | 51 | 3 | 2 | 1 | 1 | 0 | 0 | 0 | 1 | 0 | 1 | 27 | 5 | 0 | 0 | 19 |
| 2017 | 7 | 116 | 92 | 62 | 0 | 55 | 58 | 8 | 0 | 1 | 0 | 1 | 1 | 0 | 1 | 2 | 2 | 25 | 8 | 2 | 0 | 22 |
| 2018 | 3 | 145 | 115 | 49 | 0 | 50 | 80 | 10 | 0 | 1 | 1 | 0 | 0 | 0 | 1 | 0 | 1 | 17 | 2 | 0 | 0 | 22 |
| 2019 | 10 | 144 | 79 | 50 | 0 | 44 | 84 | 6 | 0 | 1 | 2 | 1 | 2 | 0 | 1 | 0 | 3 | 18 | 4 | 1 | 0 | 19 |
| 2020 | 0 | 8 | 4 | 0 | 0 | 8 | 4 | 0 | 0 | 0 | 0 | 4 | 0 | 0 | 0 | 0 | 0 | 0 | 0 | 0 | 0 | 2 |
| 9999 | 0 | 1 | 0 | 1 | 0 | 0 | 0 | 0 | 0 | 0 | 0 | 0 | 0 | 0 | 0 | 0 | 0 | 4 | 0 | 0 | 0 | 5 |

Table 2: The contents of the *hep-ex* repository, after selecting all articles from the *arXiv-with-journal* set (after Step 3), interpreting that journal reference and identifying the journal and the year of publication in the journal (equal to year in Table). Some more explanations are in the text. The abbreviations for the journal names are explained in Appendix A.

**Comparison of PLB and the arXiv.**

We will proceed by comparing the arXiv to PLB first. The main reason is that in principle all articles published in PLB after ~1992 should also be on the arXiv, if authors follow the path outlined above: submit manuscript to be published to ArXiv and journal for refereeing and once article is published in journal, the arXiv information is updated to include the journal reference.

The next step is to count the articles in PLB over the years. This exercise was done in a previous publication [4], where all PLB articles published in PLB until the end of 2019 were counted. This information can be found in Table 1 in that publication and it is explained there which PLB publications are included in the counts. Here the same criteria are used and that data set is called *PLB-pubs-all*. The Table in [4] lists all PLB publications since 1967. For this report only the years, for which possibly arXiv information exists, will be used.

In principle all arXiv repositories considered here can contribute articles to PLB. So to be able to compare the two, the ArXiv data set *ArXiv-with-journal* is split into two new data sets, one with only arXiv articles that have a PLB journal reference (*ArXiv-with-PLB-only*) and one with all arXiv articles with other (non PLB) journal references (*arXiv-with-non-PLB*). Simply counting the number of articles per year in *arXiv-with-non-PLB*, results in the number of ArXiv articles with a PLB journal reference in a given year, where the year is "year published in PLB". Table 3 shows these results.

| Year | Total PLB | Total arXiv | | Year | Total PLB | Total arXiv | | Year | Total PLB | Total arXiv | | Year | Total PLB | Total arXiv |
|---|---|---|---|---|---|---|---|---|---|---|---|---|---|---|
| 1990 | 1686 | 3 | | 1998 | 1766 | 1399 | | 2006 | 1000 | 858 | | 2014 | 816 | 347 |
| 1991 | 1742 | 26 | | 1999 | 1436 | 1171 | | 2007 | 842 | 700 | | 2015 | 827 | 278 |
| 1992 | 1721 | 362 | | 2000 | 1382 | 1191 | | 2008 | 929 | 790 | | 2016 | 880 | 294 |
| 1993 | 1617 | 696 | | 2001 | 1335 | 1189 | | 2009 | 930 | 710 | | 2017 | 877 | 303 |
| 1994 | 1485 | 868 | | 2002 | 1155 | 1009 | | 2010 | 769 | 603 | | 2018 | 849 | 264 |
| 1995 | 1571 | 1058 | | 2003 | 968 | 865 | | 2011 | 1010 | 655 | | 2019 | 807 | 228 |
| 1996 | 1653 | 1225 | | 2004 | 1038 | 935 | | 2012 | 869 | 336 | | | | |
| 1997 | 1546 | 1171 | | 2005 | 955 | 836 | | 2013 | 779 | 331 | | **Sums** | **35240** | **20701** |

Table 3: For each year (year published in PLB) the columns labeled *Total PLB* display the number of articles that year published in PLB, as explained in text. The columns labeled *Total arXiv* contains the number of articles on the arXiv (in all repositories selected) with a journal reference to PLB in that year. The "Sums" are the total number of articles in the two columns.

The results in Table 3 are a bit surprising. The expectation was that most articles in PLB would have a reference on the arXiv. That is not the case for most years. To see this graphically Figure 1 below shows the fraction of the PLB articles in each year with a journal reference on the arXiv. This is simply the ratio of the columns: *Total arXiv/Total PLB*. In an ideal world, all PLB publications should correspond to an article on arXiv with the PLB reference in it.

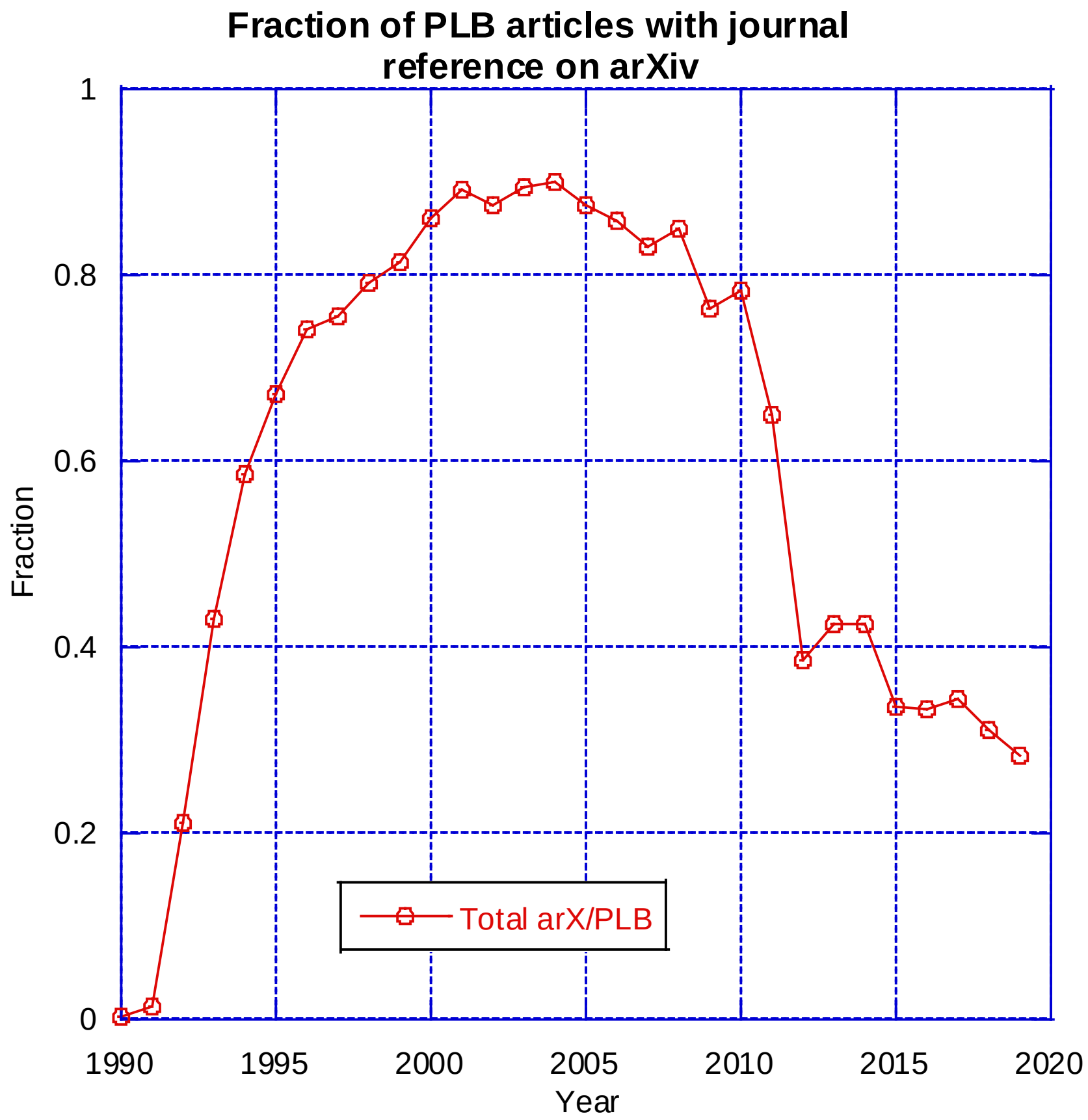

Figure 1: Fraction of PLB articles in a given year that have an arXiv article with a reference to PLB.

*Some observations:*

Initially when the arXiv was ramping up ( 1992 through 1995 with some repositories starting later) referencing the journal published version of the article increased and nearly reached unity. This would be an ideal state. Starting in 2007 we see the start of a decline in referencing, which continues until about 2011 and then flattens at about 40%. It then declines even more. This was very surprising and seeing this result for the first time, an obvious conclusion was that there must be a mistake in identifying PLB references correctly. Quite a bit of time was spent on verifying that indeed nearly all of the PLB references, in their different forms and shapes, were found. After a lot cross checks, it is felt that this information is indeed correct. Of course these incomplete data on the arXiv makes it impossible to use the arXiv for the purpose outlined at the beginning of this document.

To further verify this finding, we want to use the subject areas in PLB to see whether there is a systematic effect in them, in terms of referencing the journal published version in the arXiv publication. The subject areas in PLB are *Astrophysics & Cosmology* (Astr), *Experiments* (Expe), *Phenomenology* (phen) and *Theory* (Theo), as explained in [4]. In order to do this, it is necessary to match each PLB reference in an arXiv publication with the actual article in PLB, to establish the correlation between the subject area in PLB and the repository on the arXiv. Technically the way this was done is to read all arXiv titles from the *ArXiv-with-PBL-only* data set and storing them, including the PLB reference. Then the PLB publications starting in year 1990 are read and if a PLB article matches the reference in an arXiv article, that information is added to the PLB publication summary file. So after this process we have a new PLB publication summary file with all publications and the ones with an arXiv reference are tagged as such. This PLB data set, still containing all PLB publications, but some with arXiv tags, is called *PLB-arX-tag*. This process again requires interpreting the PLB reference correctly on the arXiv and actually matching it to a real PLB publication. The total number of PLB articles (after 1990) is 35240. The total number of arXiv articles with a PLB reference is 20829. Of those 128 have a PLB reference that cannot be interpreted (either incomplete, wrong reference, mistyped, etc.), cannot be fixed and therefore does not match to a real PLB publication. So in the end we end up with 20701 PLB publication with a reference on the arXiv. The *PLB-arX-tag* data set is now analyzed and for each publication the PLB subject area is assigned, as well as the arXiv repository if it exists. This results in "correlation matrices" for each year which show how articles move from the arXiv to PLB. Two examples for the years 2005 (high referencing on arXiv) and 2012 (low referencing on arXiv ) are given below. These years were selected because PLB had completely implemented the subject areas by 2005.

| | | Astr | Expe | Phen | Theo | **Total** |
|---|---|---|---|---|---|---|
| 2005 | hep-ex | | 84 | 1 | | **85** |
| 2005 | hep-ph | 29 | 8 | 310 | 12 | **359** |
| 2005 | hep-th | 38 | | 16 | 199 | **253** |
| 2005 | hep-lat | | | 26 | 6 | **32** |
| 2005 | nucl-ex | | 15 | | | **15** |
| 2005 | nucl-th | 1 | | 51 | | **52** |
| 2005 | astro-pl | 33 | 4 | 3 | | **40** |
| 2005 | **Sums** | **101** | **111** | **407** | **217** | **836** |

| | | Astr | Expe | Phen | Theo | **Total** |
|---|---|---|---|---|---|---|
| 2012 | hep-ex | | 94 | | | **94** |
| 2012 | hep-ph | 7 | 2 | 93 | 6 | **108** |
| 2012 | hep-th | 9 | | | 52 | **61** |
| 2012 | hep-lat | | | 2 | 10 | **12** |
| 2012 | nucl-ex | | 14 | | | **14** |
| 2012 | nucl-th | | | 8 | 16 | **24** |
| 2012 | astro-pl | 19 | 1 | 2 | 1 | **23** |
| 2012 | **Sums** | **35** | **111** | **105** | **85** | **336** |

Table 4: The correlation matrix for the two years 2005 and 2012. It shows how many articles from a given arXiv repository (*hep-ex, hep-ph, hep-th, hep-lat, nucl-ex, nucl-th and astro-ph*) end up in each PLB subject area (*Astr, Expe, Phen, Theo*). Of course only PLB publications with a reference on the arXiv are counted.

These "correlation matrices" for each year show how articles flow. An example for 2005: A total of 407 articles were published in PLB in the subject area *Phen*. Of those 1 came from *hep-ex*, 310 from *hep-ph*, 16 from *hep-th*, 26 from *hep-lat*, 51 from *nucl-th* and 3 from *astro-ph*. One can also read it the other way: how *hep-ph* contributes to PLB subject areas. Of course the numbers only include PLB publications that have a reference on the arXiv. By using the "Sums" row, one can now see how many articles in each PLB subject area are referenced on the arXiv. This information is displayed in Table 5. Only the years for which subject areas exist are displayed.

| Year | Astro PLB | Expe PLB | Pheno PLB | Theory PLB | Astro arX | Expe arX | Pheno arX | Theory arX |
|---|---|---|---|---|---|---|---|---|
| 2004 | 12 | 17 | 65 | 46 | 11 | 13 | 58 | 38 |
| 2005 | 124 | 152 | 426 | 253 | 101 | 111 | 407 | 217 |
| 2006 | 158 | 135 | 444 | 263 | 124 | 100 | 419 | 215 |
| 2007 | 89 | 105 | 385 | 263 | 43 | 72 | 368 | 217 |
| 2008 | 134 | 100 | 404 | 291 | 95 | 69 | 385 | 241 |
| 2009 | 125 | 115 | 362 | 328 | 69 | 75 | 321 | 245 |
| 2010 | 133 | 94 | 293 | 249 | 70 | 63 | 272 | 198 |
| 2011 | 124 | 145 | 427 | 314 | 57 | 84 | 323 | 191 |
| 2012 | 106 | 168 | 319 | 276 | 35 | 111 | 105 | 85 |
| 2013 | 76 | 140 | 282 | 281 | 30 | 93 | 111 | 97 |
| 2014 | 113 | 114 | 299 | 290 | 50 | 64 | 121 | 112 |
| 2015 | 88 | 133 | 295 | 311 | 25 | 63 | 98 | 92 |
| 2016 | 85 | 147 | 358 | 290 | 18 | 86 | 97 | 93 |
| 2017 | 88 | 141 | 337 | 311 | 18 | 68 | 117 | 100 |
| 2018 | 68 | 142 | 313 | 326 | 14 | 70 | 95 | 85 |
| 2019 | 87 | 126 | 267 | 327 | 26 | 69 | 68 | 65 |

Table 5: The number of publications in PLB in a given subject area per year (with label "PLB") and the corresponding number of arXiv articles with a matching reference to an article in the PLB subject area (with label "arX"). Example: In 2005, out of 426 articles published in PLB under Phenomenology, 407 had an arXiv article with a matching reference.

Figure 2 shows the fraction of PLB articles in each subject area with a matching arXiv article. There are no data before 2004, because subject areas started in PLB in 2004.

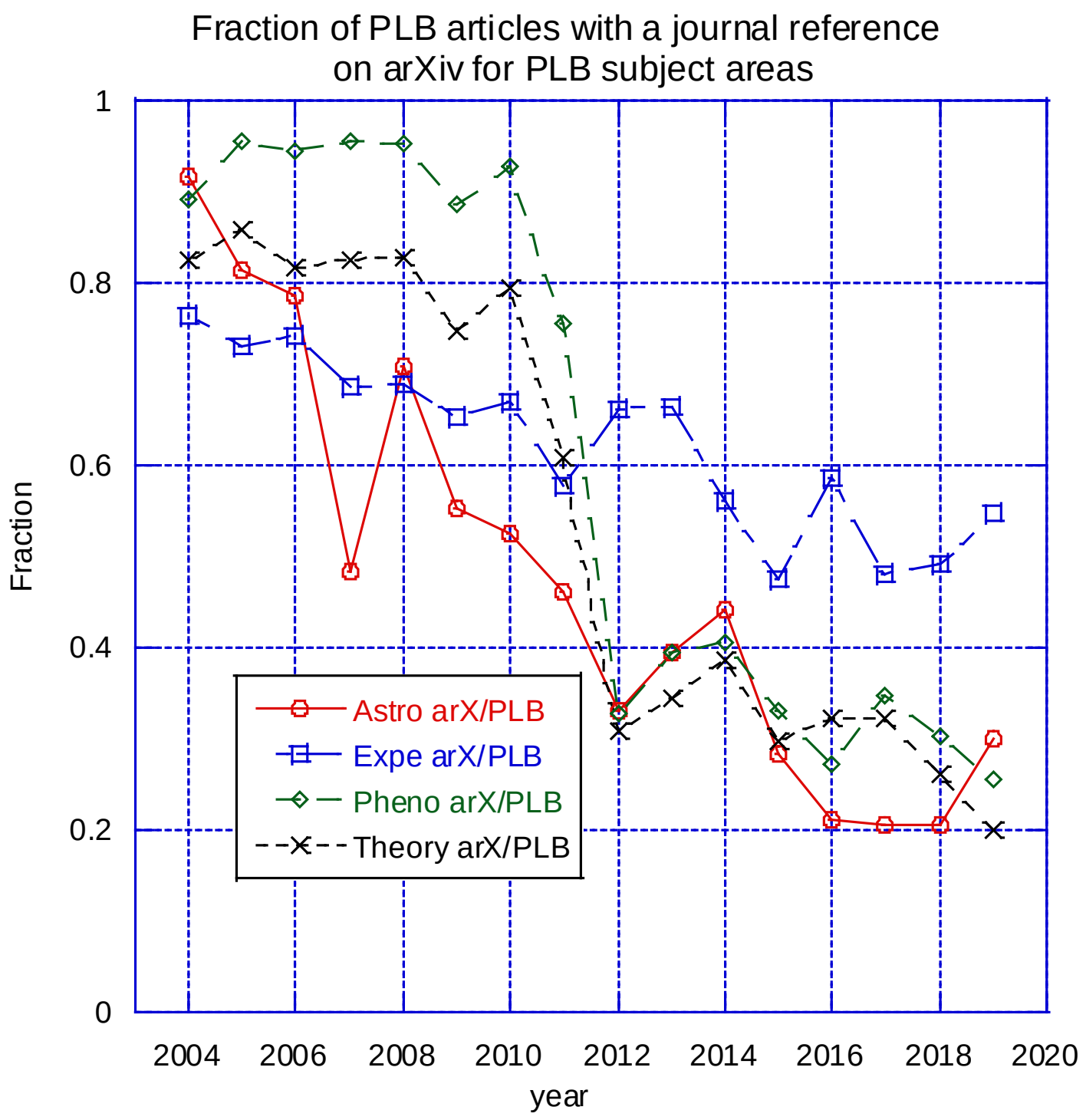

Figure 2: Same information as in Figure 1, but now showing it for the four subject areas in PLB.

The above figure basically shows the same behavior as Figure 1. Starting around 2005 there is a steep decline in the fraction of PLB publications with a journal reference in an arXiv article for all subject areas. The individual subject areas in PLB follow the same trend as the overall publications. The only subject area that seems to be doing a bit better is *Experiments (Expe)*.

As mentioned above, when seeing this information for the first time the suspicion was that something was wrong and many checks were done. There are many more articles (in later years) in PLB than articles on the arXiv with a PLB reference. The question was: are these articles or the arXiv and the reference to PLB is missing or are they not on the arXiv? For the following we simply call these PLB articles "*missing PLB articles on arXiv*"

To address this question it was decided to try and look for the *missing PLB articles on arXiv*. This was pursued through matching the titles of the PLB publications without an arXiv reference with all possible arXiv titles. It might have been better to try and match authors or title and authors, but that was not pursued. Some time was spent on finding existing algorithms that can do title matching, but nothing simple was found. After spending quite, a bit of time trying things, a simple algorithm was used, that does the following:

- Only compare a PLB title to a possible arXiv title, if the arXiv article is in the same or previous year as the PLB publication. This is mostly for speeding up the matching and minimizing wrong matches.
- Ignore all words from both titles that have only one letter.
- Ignore all words from both titles that have less than 4 characters, except if they are all capital.
- Ignore all words that contain a $ sign (arXiv allows TeX symbols in titles, PLB does not).
- Count the number of words in two titles to be matched, irrespective of their location in the title.
- Construct the ratios: *RP = #matched words/#words-in-PLB-title* and *RA = #matched words/#words- in-arXiv-title*.
- Require both *RA* and *RP* to be equal to 1.0 for a match.

One can allow smaller values for *RA* and *RP* (many were tried), and optimize the efficiency of identifying all the right matches. However, in this case many matches are done and it is important to reject random matches of titles. The algorithm was tested by using the PLB-arX-tag sample, which contains all PLB publication titles and, if it exists, the ArXiv article information including the arXiv title. These two titles should be the same. The efficiency of the algorithm is obtained by calculating *RA* and *RP* for the PLB title and the arXiv title that contains the matching PLB reference. The matching efficiency is then the fraction of titles found for a given value of *RA* and *RP*. This efficiency is shown in Table 3a is a function of the parameters *RA* and *RP*.

| RA to right | | | | | | | | | | | | |
|---|---|---|---|---|---|---|---|---|---|---|---|---|
| RP down | | 1 | 0.9 | 0.8 | 0.7 | 0.6 | 0.5 | 0.4 | 0.3 | 0.2 | 0.1 | 0 |
| 1 | | 0.64 | 0.65 | 0.67 | 0.67 | 0.68 | 0.68 | 0.68 | 0.68 | 0.68 | 0.68 | 0.68 |
| 0.9 | | 0.65 | 0.67 | 0.70 | 0.70 | 0.71 | 0.71 | 0.71 | 0.71 | 0.71 | 0.71 | 0.71 |
| 0.8 | | 0.69 | 0.71 | 0.82 | 0.86 | 0.88 | 0.89 | 0.89 | 0.89 | 0.89 | 0.89 | 0.89 |
| 0.7 | | 0.70 | 0.72 | 0.83 | 0.90 | 0.93 | 0.94 | 0.95 | 0.95 | 0.95 | 0.95 | 0.95 |
| 0.6 | | 0.71 | 0.73 | 0.84 | 0.91 | 0.96 | 0.97 | 0.98 | 0.98 | 0.98 | 0.98 | 0.98 |
| 0.5 | | 0.71 | 0.73 | 0.85 | 0.91 | 0.96 | 0.98 | 0.99 | 0.99 | 0.99 | 0.99 | 0.99 |
| 0.4 | | 0.71 | 0.73 | 0.85 | 0.91 | 0.96 | 0.98 | 0.99 | 0.99 | 0.99 | 0.99 | 0.99 |
| 0.3 | | 0.71 | 0.73 | 0.85 | 0.91 | 0.96 | 0.98 | 0.99 | 0.99 | 1.00 | 1.00 | 1.00 |
| 0.2 | | 0.71 | 0.73 | 0.85 | 0.91 | 0.96 | 0.98 | 0.99 | 1.00 | 1.00 | 1.00 | 1.00 |
| 0.1 | | 0.71 | 0.73 | 0.85 | 0.91 | 0.96 | 0.98 | 0.99 | 1.00 | 1.00 | 1.00 | 1.00 |
| 0 | | 0.71 | 0.73 | 0.85 | 0.91 | 0.96 | 0.98 | 0.99 | 1.00 | 1.00 | 1.00 | 1.00 |

Table 6a: Efficiency of title match as function of *RA* and *RP*.

| RA to right | | | | | | | | | | | | |
|---|---|---|---|---|---|---|---|---|---|---|---|---|
| RP down | | 1 | 0.9 | 0.8 | 0.7 | 0.6 | 0.5 | 0.4 | 0.3 | 0.2 | 0.1 | 0 |
| 1 | | 4.8E-05 | 4.8E-05 | 4.8E-05 | 4.8E-05 | 4.8E-05 | 4.8E-05 | 4.8E-05 | 4.8E-05 | 4.8E-05 | 4.8E-05 | 4.8E-05 |
| 0.9 | | 4.8E-05 | 4.8E-05 | 4.8E-05 | 4.8E-05 | 4.8E-05 | 4.8E-05 | 4.8E-05 | 4.8E-05 | 4.8E-05 | 4.8E-05 | 4.8E-05 |
| 0.8 | | 4.8E-05 | 9.7E-05 | 2.9E-04 | 3.4E-04 | 4.3E-04 | 4.3E-04 | 5.3E-04 | 5.3E-04 | 5.3E-04 | 5.3E-04 | 5.3E-04 |
| 0.7 | | 4.8E-05 | 9.7E-05 | 2.9E-04 | 3.9E-04 | 6.3E-04 | 7.2E-04 | 8.7E-04 | 9.7E-04 | 1.0E-03 | 1.0E-03 | 1.0E-03 |
| 0.6 | | 9.7E-05 | 1.4E-04 | 3.4E-04 | 5.3E-04 | 9.7E-04 | 1.4E-03 | 1.7E-03 | 2.2E-03 | 2.7E-03 | 2.7E-03 | 2.7E-03 |
| 0.5 | | 1.4E-04 | 1.9E-04 | 5.3E-04 | 7.7E-04 | 1.4E-03 | 2.7E-03 | 3.8E-03 | 5.1E-03 | 6.6E-03 | 7.2E-03 | 7.2E-03 |
| 0.4 | | 1.4E-04 | 1.9E-04 | 5.3E-04 | 8.2E-04 | 1.9E-03 | 3.8E-03 | 5.6E-03 | 8.0E-03 | 1.2E-02 | 1.2E-02 | 1.3E-02 |
| 0.3 | | 1.4E-04 | 1.9E-04 | 5.8E-04 | 9.7E-04 | 2.5E-03 | 5.1E-03 | 8.2E-03 | 1.3E-02 | 2.0E-02 | 2.4E-02 | 2.4E-02 |
| 0.2 | | 1.4E-04 | 1.9E-04 | 5.8E-04 | 9.7E-04 | 2.6E-03 | 6.7E-03 | 1.2E-02 | 2.0E-02 | 4.2E-02 | 6.9E-02 | 7.2E-02 |
| 0.1 | | 1.9E-04 | 2.4E-04 | 6.3E-04 | 1.0E-03 | 2.7E-03 | 7.2E-03 | 1.3E-02 | 2.4E-02 | 6.7E-02 | 1.5E-01 | 1.7E-01 |
| 0 | | 1.9E-04 | 2.4E-04 | 6.3E-04 | 1.0E-03 | 2.7E-03 | 7.2E-03 | 1.3E-02 | 2.4E-02 | 7.0E-02 | 1.6E-01 | 1.0E+00 |

Table 6b: Probability(*Pran*) to find a wrong title match. *Rejection = (1 – Pran)*

Table 3a shows the efficiency with which the algorithm finds the correct title, if it is in the sample considered. This efficiency is given as a function of the parameters *RA* and *RP*. Obviously this efficiency approaches 1, as the parameters go down. Choosing one parameter to be zero shows the efficiency of the other parameter used by itself. The bottom line is that with our choice of requiring *RA = RP = 1*, the efficiency is 64% for finding the right title. Table 6b shows the probability (*Pran*) to find a wrong title match. This is typically expressed as rejection, where rejection is (1- *Pran)*. So in this case we see the rejection is very high for *RA= RP = 1* and it is 99.99995%. This probability is obtained by using the same sample as for the efficiency, but comparing the arXiv title to the title of the PLB article that is published next in PLB. Those two titles should not match, so finding a match is the probability for a wrong match. For both Tables the data set used contained 20700 titles.

This title match algorithm is applied to the *missing PLB articles on arXiv* (14539 publications). The PLB title, without an arXiv tag, is compared to all arXiv titles in the *arXiv-without-journal* dataset. This dataset contains a total of 249174 arXiv titles from all repositories considered. For each PLB title a title match is performed, with the algorithm outlined above, and a total of 2190 matches are found. This corresponds to 1.9 $10^8$ attempted title matches.

Before looking at these matched titles it is important to check how many random title matches there are. To do this, the same PLB data set is used (ones without an arXiv reference, 14536), but now these titles are matched to the *arXiv-with-journal* dataset. This dataset, with 254472 titles, also includes all arXiv titles with a reference to PLB. A total of 1.6 $10^8$ title matches are performed and 89 matches are found. In an ideal world one expects zero. All of these 89 matches were scanned and in 78 of them some form of a reference to PLB was found. However, the reference is incomplete, has typos, missing information, wrong page number etc. So these are PLB references on the arXiv, which the algorithm interpreting the journal reference on the arXiv, could not convert to a correct reference to an existing PLB publication. These exist in the *arXiv-with-journal* data set and there are 128 of them (see above). Given the efficiency of the title matching algorithm, we expect to find 128 x 0.64 = 81.9 matches. This is in very good agreement with the 78 found. Eleven out of the 89 title matches found were references to other journals (random matches). This number is very small compared to the number of matches tried, so the conclusion is that the sample of 2190 matches above is very pure and these are indeed PLB publications.

Returning to the 2190 title matches found, they are analyzed the same way as was done above for the PLB publications with a reference on the arXiv. The correlation matrices are created for each year, resulting in the information between the PLB subject areas and the arXiv repositories, where they came from. This information is used to create a new updated table, including information already shown in Tables 3 and 5. This is Table 7 shown below. All information is shown for the year in which the PLB articles are published. The blue area of Table 7 contains the information about PLB publications in each year. In 2004 the previously mentioned subject areas started. That is why before 2004 all publications are counted under the column “None” i.e. there is no subject are. The columns shaded green contain the number of articles with a journal reference found on the arXiv, subdivided into the four PLB subject areas, if they exist (after 2004). All this information was already in Table 3 and 5, but it is put together here to be comprehensive. The new information added, shaded in yellow, are the results of the title matching procedure described above and resulting in 2190 articles found on the arXiv, that correspond to a PLB publication. In the table this total is broken down by year and by subject areas, when they exist. With the information in this table it is now possible to calculate some interesting fractions and show the results graphically and compare to previous results.

| Year | Astr PLB | Expe PLB | Phen PLB | Theo PLB | None PLB | Total PLB | Astr arX | Expe arX | Phen arX | Theo arX | None arX | Total arX | Astr-Mat | Expe-Mat | Phen-Mat | Theo-Mat | None-Mat | Total-Mat |
|---|---|---|---|---|---|---|---|---|---|---|---|---|---|---|---|---|---|---|
| **1990** | | | | | 1686 | 1686 | | | | | 3 | 3 | 0 | 0 | 0 | 0 | 0 | 0 |
| **1991** | | | | | 1742 | 1742 | | | | | 26 | 26 | 0 | 0 | 0 | 0 | 0 | 0 |
| **1992** | | | | | 1721 | 1721 | | | | | 362 | 362 | 0 | 0 | 0 | 0 | 1 | 1 |
| **1993** | | | | | 1617 | 1617 | | | | | 696 | 696 | 0 | 0 | 0 | 0 | 6 | 6 |
| **1994** | | | | | 1485 | 1485 | | | | | 868 | 868 | 0 | 0 | 0 | 0 | 3 | 3 |
| **1995** | | | | | 1571 | 1571 | | | | | 1058 | 1058 | 0 | 0 | 0 | 0 | 7 | 7 |
| **1996** | | | | | 1653 | 1653 | | | | | 1225 | 1225 | 0 | 0 | 0 | 0 | 4 | 4 |
| **1997** | | | | | 1546 | 1546 | | | | | 1171 | 1171 | 0 | 0 | 0 | 0 | 3 | 3 |
| **1998** | | | | | 1766 | 1766 | | | | | 1399 | 1399 | 0 | 0 | 0 | 0 | 8 | 8 |
| **1999** | | | | | 1436 | 1436 | | | | | 1171 | 1171 | 0 | 0 | 0 | 0 | 5 | 5 |
| **2000** | | | | | 1382 | 1382 | | | | | 1191 | 1191 | 0 | 0 | 0 | 0 | 4 | 4 |
| **2001** | | | | | 1335 | 1335 | | | | | 1189 | 1189 | 0 | 0 | 0 | 0 | 6 | 6 |
| **2002** | | | | | 1155 | 1155 | | | | | 1009 | 1009 | 0 | 0 | 0 | 0 | 4 | 4 |
| **2003** | | | | | 968 | 968 | | | | | 865 | 865 | 0 | 0 | 0 | 0 | 2 | 2 |
| **2004** | 12 | 17 | 65 | 46 | 898 | 1038 | 11 | 13 | 58 | 38 | 815 | 935 | 0 | 0 | 3 | 0 | 1 | 4 |
| **2005** | 124 | 152 | 426 | 253 | | 955 | 101 | 111 | 407 | 217 | | 836 | 0 | 0 | 1 | 0 | 0 | 1 |
| **2006** | 158 | 135 | 444 | 263 | | 1000 | 124 | 100 | 419 | 215 | | 858 | 0 | 0 | 2 | 1 | 0 | 3 |
| **2007** | 89 | 105 | 385 | 263 | | 842 | 43 | 72 | 368 | 217 | | 700 | 0 | 0 | 3 | 3 | 0 | 6 |
| **2008** | 134 | 100 | 404 | 291 | | 929 | 95 | 69 | 385 | 241 | | 790 | 0 | 0 | 4 | 2 | 0 | 6 |
| **2009** | 125 | 115 | 362 | 328 | | 930 | 69 | 75 | 321 | 245 | | 710 | 4 | 0 | 14 | 11 | 0 | 29 |
| **2010** | 133 | 94 | 293 | 249 | | 769 | 70 | 63 | 272 | 198 | | 603 | 3 | 2 | 5 | 7 | 0 | 17 |
| **2011** | 124 | 145 | 427 | 314 | | 1010 | 57 | 84 | 323 | 191 | | 655 | 11 | 4 | 53 | 38 | 0 | 106 |
| **2012** | 106 | 168 | 319 | 276 | | 869 | 35 | 111 | 105 | 85 | | 336 | 14 | 14 | 138 | 89 | 0 | 255 |
| **2013** | 76 | 140 | 282 | 281 | | 779 | 30 | 93 | 111 | 97 | | 331 | 20 | 7 | 101 | 76 | 0 | 204 |
| **2014** | 113 | 114 | 299 | 290 | | 816 | 50 | 64 | 121 | 112 | | 347 | 22 | 6 | 99 | 76 | 0 | 203 |
| **2015** | 88 | 133 | 295 | 311 | | 827 | 25 | 63 | 98 | 92 | | 278 | 20 | 9 | 117 | 92 | 0 | 238 |
| **2016** | 85 | 147 | 358 | 290 | | 880 | 18 | 86 | 97 | 93 | | 294 | 21 | 11 | 159 | 91 | 0 | 282 |
| **2017** | 88 | 141 | 337 | 311 | | 877 | 18 | 68 | 117 | 100 | | 303 | 22 | 10 | 133 | 92 | 0 | 257 |
| **2018** | 68 | 142 | 313 | 326 | | 849 | 14 | 70 | 95 | 85 | | 264 | 15 | 16 | 136 | 96 | 0 | 263 |
| **2019** | 87 | 126 | 267 | 327 | | 807 | 26 | 69 | 68 | 65 | | 228 | 16 | 8 | 126 | 113 | 0 | 263 |
| **Sums** | **1610** | **1974** | **5276** | **4419** | **21961** | **35240** | **786** | **1211** | **3365** | **2291** | **13048** | **20701** | **168** | **87** | **1094** | **787** | **54** | **2190** |

Table 7: Shown are the number of articles found in PLB per year (in blue section) and subdivided into the PLB subject areas, for the years when those areas exist. If they do not exist, articles are counted under "None". The "Total " is the sum of all PLB articles. The green shaded area shows the results for articles on the arXiv with a found reference to a PLB publication. Both these results were already partially shown in Tables 3 and 5. The yellow shaded area shows the number of articles found on the arXiv, corresponding to a PLB publication and found through title matching as described above. More in the text.

Following are a set of figures showing the results from Table 7 graphically. The names of the columns in Table 7 are used to describe the ratios/fractions shown in the figures.

- Figure 3 is the ratios of *Ra=Total arX / Total PLB* and *Rm =Total-Mat/Total PLB. Ra* was already shown in Figure 1 and is simply repeated here. The result clearly shows that the articles found by title matching are in the years when journal references on the arXiv were missing.
- Figure 4 shows the ratio *Rm*, but now broken down into the four subject areas. For example, for *Phen*, *Rm = Phen-Mat/Phen PLB*. This was already shown in Figure 2 before. Here the figure is repeated, but the horizontal axis is changed to match the other figures.

- Figure 5: Given the numbers in Table 7, we can estimate how many PLB articles have a corresponding article on the arXiv. This projected number of PLB articles (*Ppwa*) is equal to the actual number of articles on the arXiv with a journal reference to PLB, plus the number of arXiv articles identified by a title match, divided by the title matching efficiency (*efftm* =64%). So for example for the subject area *Theo*: *Ppwa = Theo arX + (Theo-Mat/efftm).* Plotted in Figure 5 is the fraction for each subject area. So for *Theo*, the graph shows: (*Ppwa/Theo PLB*). In an ideal world all these fractions would be one i.e. every PLB article has a corresponding article on the arXiv.
- Finally Figure 6 shows the same quantity as Figure 5, but in this case for all articles, and it extends over the whole period covered by the arXiv. In an ideal world this fraction would approach 1, after the initial ramp up of the arXiv.

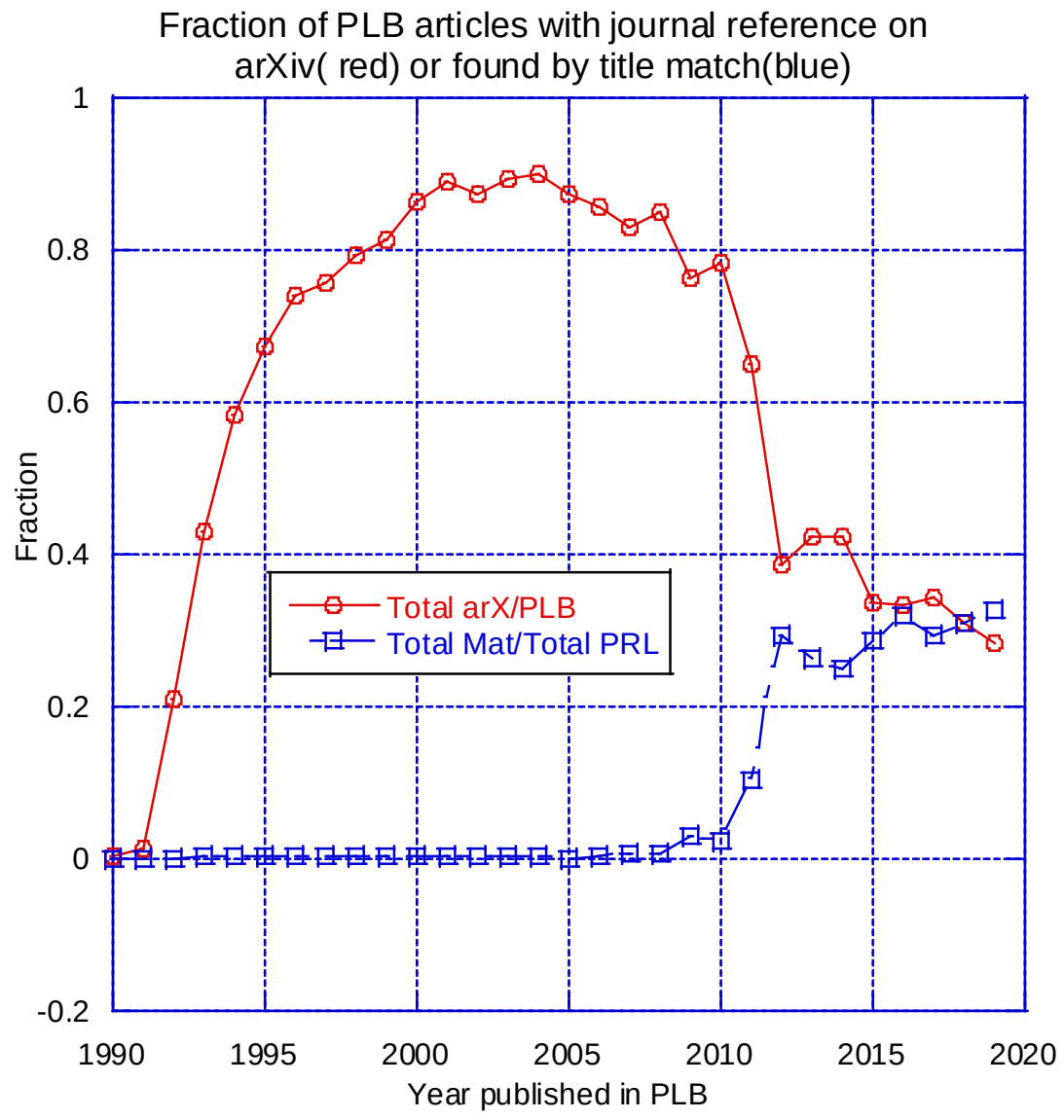

Figure 3: Fraction of PLB articles with journal reference, found on arXiv (red) or by title match (blue)

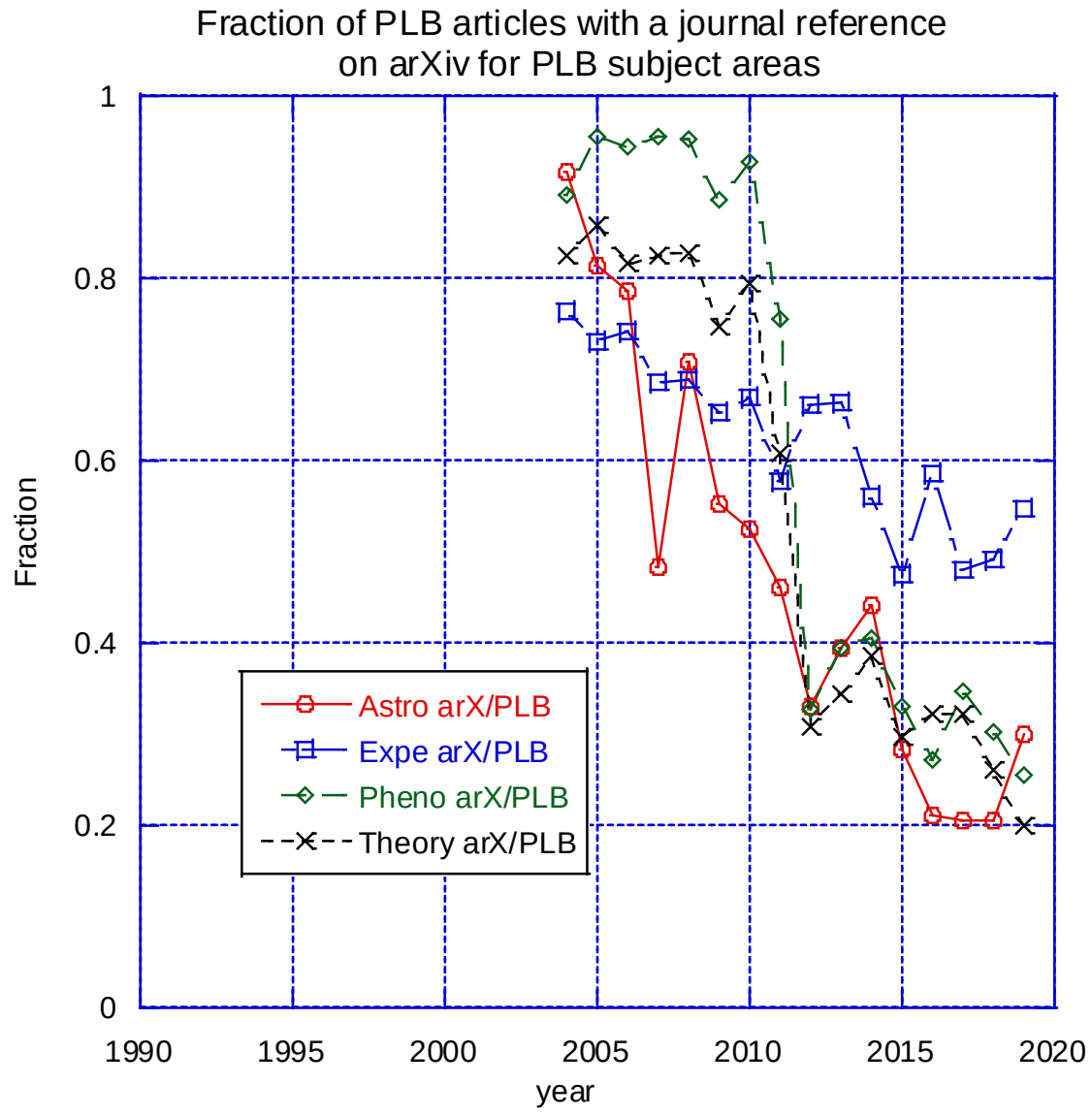

Figure 4: Same as Figure 3, but now for the different PLB subject areas

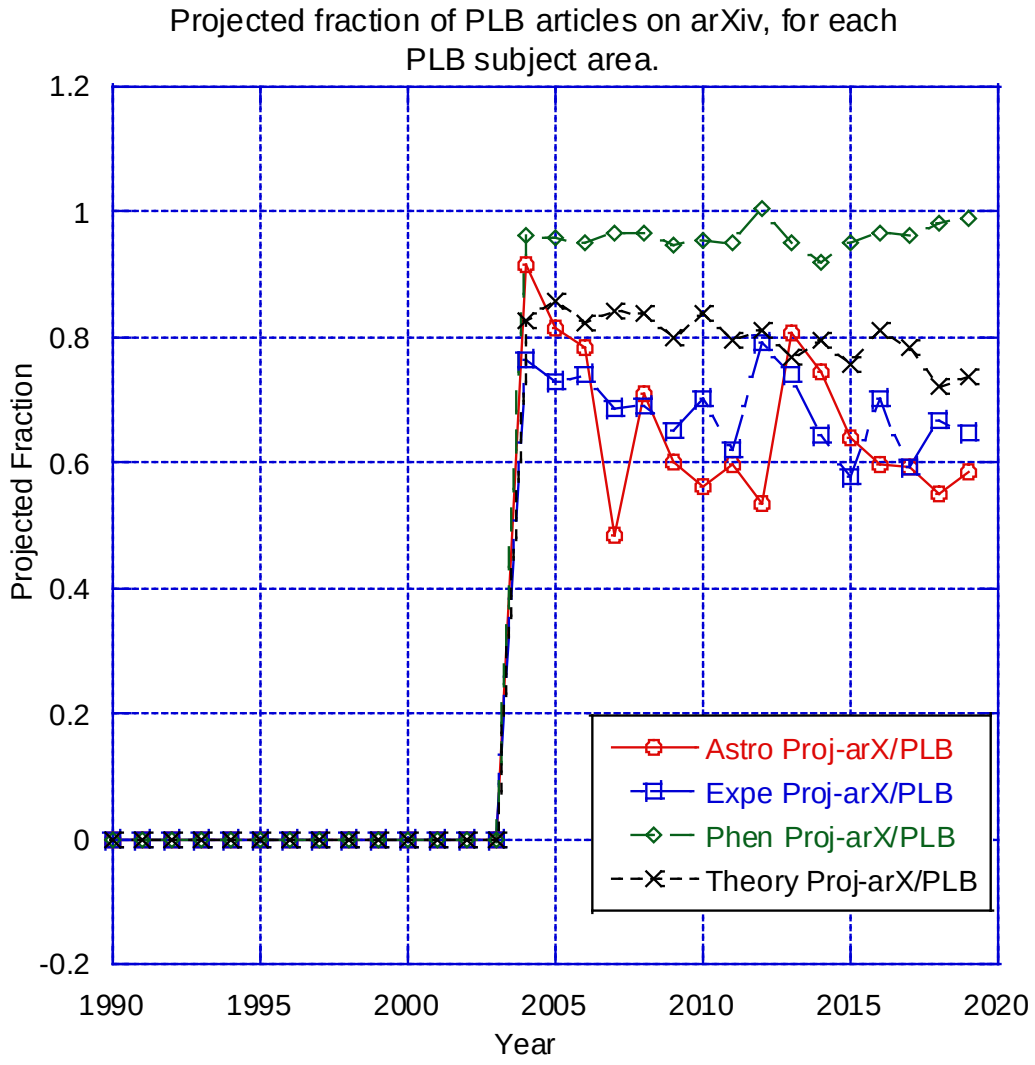

Figure 5: Projected fraction of PLB articles with corresponding arXiv article for subject areas ( see text for more detail)

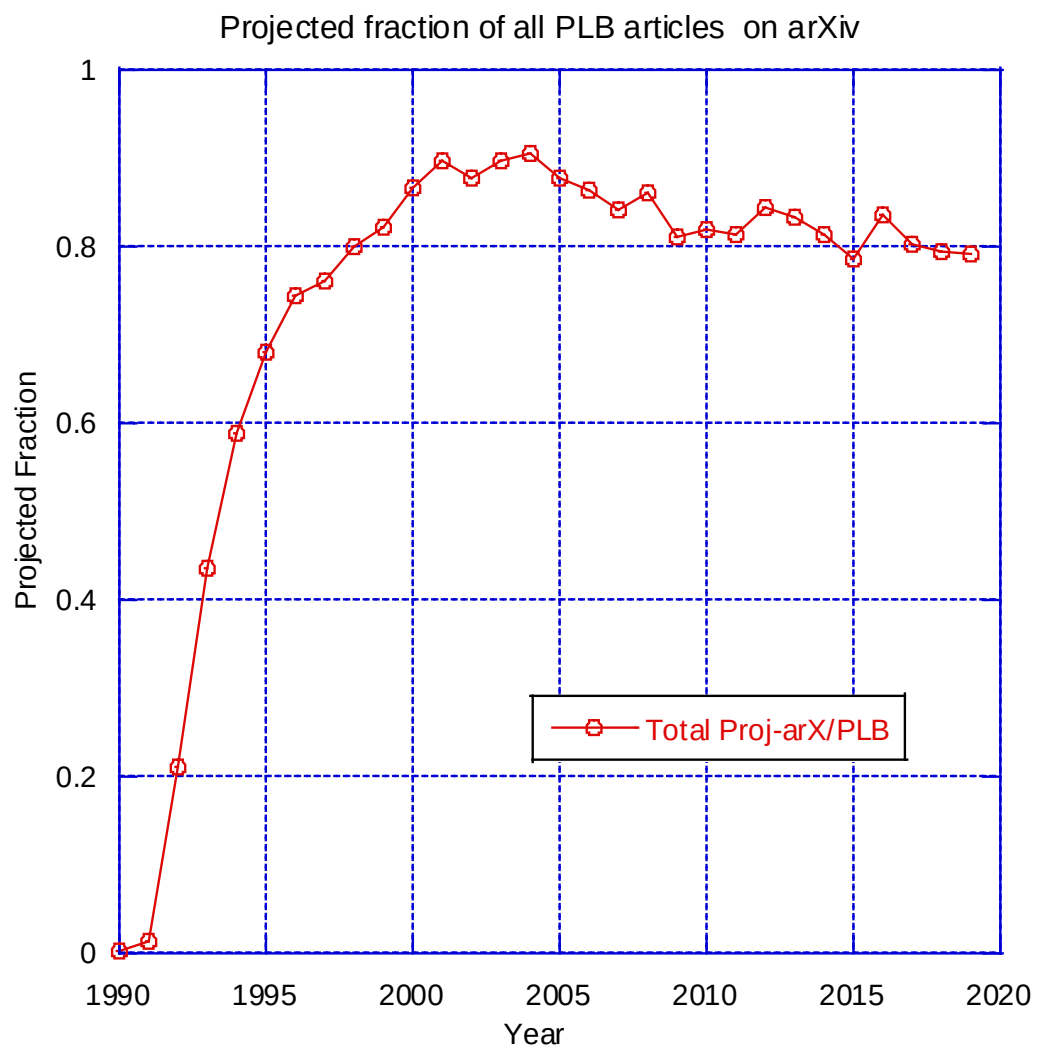

Figure 6: Same as Figure 5, but now for all PLB articles and extending for all years covered by arXiv.

*Observations and conclusions.*

The following conclusions can be drawn from these numbers and graphs:

1. Figure 3 and 6 show the ramp up of using the arXiv as the preprint server for articles published in PLB. Since the start of the arXiv in the early 90's, it took about 10 years for all communities, publishing in PLB, to start using it fully.
2. When trying to associate articles in PLB with their corresponding "preprint" version on the arXiv, the journal reference exists in most cases up to the year 2010. Since then, this reference is missing in about 60% of the cases.
3. Looking at the subject areas in PLB, basically the same conclusion can be drawn, although there are minor differences for each area.
4. This leaves one with the question; are the "preprint" articles missing from the arXiv or is the journal reference simply not made available. To answer that a title matching algorithm was developed and used to look for the missing articles on the arXiv.
5. Figures 3 shows that title matching finds articles in the time frame where journal references are missing (beyond 2010). Taking into account the efficiency of the title matching, one can calculate the total number of PLB articles with a corresponding article on the arXiv. Figure 5 & 6 show that for the subject areas and the total number of articles. The subject area *Phen* does better than all others.
6. Figure 5 shows that the subject areas *Astro* and *Expe*, have at most about 80% of their articles estimated to be on the arXiv. It is not clear what the cause is. It could simply be that articles are in arXiv repositories not considered here.
7. Although the results in Figures 5 and 6 are not perfect, and they are projections with uncertainties, the conclusion is that all (or nearly all) PLB publications have a corresponding article on the arXiv, but it is not always reflected in the journal reference.

As already stated earlier the initial finding was a surprise, but it is reassuring to see that in principle every publication in PLB has a corresponding preprint on the arXiv. The question remaining is: is this a general feature for all publications or is it journal specific? To answer this question, the above analysis was repeated for the journal Physical Review Letters. This is described in the next chapter.

## Comparison of PRL and the arXiv

Physical Review Letters (PRL) serves the whole physics community and has many subject areas. To compare to the study described above, three subject areas were selected from all PRL volumes over the years. They are *Gravitation and Astrophysics*(GA), *Elementary Particles & Fields*(EPF) and *Nuclear Physics*(NP). Appendix C gives a brief introduction and history of PRL and on how and what information from PRL is used in this report. The Appendix has information for all years for PRL. The data set containing all PRL articles in all subject areas is called *PRL-pubs-all*. For the remainder of this report we will only consider the years after 1990. The three PRL subject areas are selected from *PRL-pubs-all*. They match closely the subject areas covered by PLB and they should basically be covered by the same arXiv repositories as were used for the PLB analysis above.

The comparison of the three selected PRL subject areas to the arXiv contents, is done in the same way as the comparison to PLB. The main difference is that for PLB all articles were used (PLB only publishes results in *Astrophysics*, *Nuclear Physics* and *Particle Physics*), whereas for PRL only the three above mentioned subject areas are used. So in principle an arXiv article, with a journal reference to PRL, could point to a subject area in PRL not considered here. The description of the steps done to get the PRL results will be much shorter and only differences to the more detailed description for PLB above, will be given.

The same arXiv data set as described above in “Approach and Data Gathering” i.e. *arXiv-with-journal* and *arXiv-without-journal* is used for this part. As done above the *arXiv-with-journal*-is split into a data set with PRL journal references (called *arXiv-with-PRL-only* with 8826 articles) and a data set with arXiv articles that contain a journal reference, but it is not to PRL (*arXiv-with-non-PRL*). As before this step requires an algorithm to correctly identify references to PRL and we will come back to that later.

As before each PRL reference in an arXiv publication is matched with the actual article in PRL to establish the correlation between the subject area in PRL and the repository on the arXiv. As input the *PRL-pubs-all* data set is used to create a new PRL data set, with arXiv tags, and it is called *PRL-arX-tag*. It should be noted that this new tagged data set still contains all PRL publications from all subject areas. This process requires interpreting the PRL reference on the arXiv and actually matching it to a real PRL publication. The total number of PRL articles (after 1990, in the *GA*, *NP* and *EPF* subject areas) is 12420 (about 1/3 of the PLB articles). The total number of arXiv articles with a PRL reference is 8849. Of those 62 have a PRL reference that cannot be interpreted (either incomplete, wrong reference, mistyped, etc.), cannot be fixed and therefore does not match to a real PRL publication. So in the end we end up with 8787 PRL publications with a matched reference on the arXiv. Some of the arXiv journal references do not point to the *GA*, *NP* or *EPF* subject areas, but to another subject area in PRL. These articles are counted under “Other” in the tables below. A total of 521 articles like that were found. The *PRL-arX-tag* dataset is now analyzed and for each publication the PRL subject area is assigned, as well as the arXiv repository if it exists. This results in the “correlation matrices” for each year which show how articles move from the arXiv to PRL. Two examples for the years 2005 (high referencing on arXiv) and 2012 (low referencing on arXiv ) are given below.

| | | EPF | NP | GA | Other | Total |
|---|---|---|---|---|---|---|
| **2005** | **hep-ex** | 119 | 3 | 1 | 1 | **124** |
| **2005** | **hep-ph** | 58 | 13 | 10 | 7 | **88** |
| **2005** | **hep-th** | 23 | | 13 | 4 | **40** |
| **2005** | **hep-lat** | 11 | 1 | | 1 | **13** |
| **2005** | **nucl-ex** | 8 | 34 | 1 | | **43** |
| **2005** | **nucl-th** | 2 | 37 | 1 | 2 | **42** |
| **2005** | **astro-ph** | | | 56 | 5 | **61** |
| **2005** | **Sums** | **221** | **88** | **82** | **20** | **411** |

| | | EPF | NP | GA | Other | Total |
|---|---|---|---|---|---|---|
| **2012** | **hep-ex** | 78 | 4 | 1 | 1 | **84** |
| **2012** | **hep-ph** | 29 | 4 | 5 | 2 | **40** |
| **2012** | **hep-th** | 13 | | 6 | 2 | **21** |
| **2012** | **hep-lat** | 8 | 1 | | 2 | **11** |
| **2012** | **nucl-ex** | 1 | 19 | | 2 | **22** |
| **2012** | **nucl-th** | 1 | 18 | 1 | | **20** |
| **2012** | **astro-ph** | | 1 | 38 | 5 | **44** |
| **2012** | **Sums** | **130** | **47** | **51** | **14** | **242** |

Table 8: The "correlation matrix" for the two years 2005 and 2012. It shows how many articles from a given arXiv repository (*hep-ex*, *hep-ph*, *hep-th*, *hep-lat*, *nucl-ex*, *nucl-th and astro-ph*) end up in each PRL subject area (*EPF*, *GA*, *NP* and *Other*). Of course only PRL publications with a reference on arXiv are counted.

Given the "correlation matrices" for each year, we can construct the Table 9 for PRL, which is the same as Table 7 above for PLB. In addition to using the "correlation matrices" based on the journal references on the arXiv, title matching is performed on those PRL publications that do not have an existing journal reference on the arXiv. This is done exactly same as for PLB, described above. The efficiency for matching is re-determined for PRL (different journals use different ways to display their titles). The title matching efficiency is found to be 51.2%.

| Year | EPF PRL | NP PRL | GA PRL | Total | EPF arX | NP arX | GA arX | All arX | # not on arX | Other arX | EPF-Mat | NP-Mat | GA-Mat | Total-Mat |
|---|---|---|---|---|---|---|---|---|---|---|---|---|---|---|
| **1990** | 171 | 85 | 62 | 318 | 0 | 0 | 0 | 0 | 318 | | 0 | 0 | 0 | 0 |
| **1991** | 133 | 100 | 61 | 294 | 2 | 0 | 0 | 2 | 292 | 1 | 0 | 0 | 0 | 0 |
| **1992** | 159 | 106 | 64 | 329 | 32 | 4 | 17 | 53 | 276 | 12 | 0 | 0 | 0 | 0 |
| **1993** | 163 | 110 | 73 | 346 | 75 | 11 | 26 | 112 | 234 | 18 | 0 | 0 | 0 | 0 |
| **1994** | 124 | 106 | 58 | 288 | 77 | 19 | 30 | 126 | 162 | 21 | 0 | 1 | 0 | 1 |
| **1995** | 155 | 131 | 81 | 367 | 112 | 24 | 37 | 173 | 194 | 24 | 0 | 2 | 0 | 2 |
| **1996** | 140 | 122 | 87 | 349 | 110 | 31 | 56 | 197 | 152 | 24 | 0 | 0 | 0 | 0 |
| **1997** | 165 | 112 | 87 | 364 | 143 | 34 | 56 | 233 | 131 | 21 | 0 | 0 | 0 | 0 |
| **1998** | 181 | 117 | 106 | 404 | 157 | 46 | 63 | 266 | 138 | 18 | 0 | 0 | 0 | 0 |
| **1999** | 157 | 161 | 82 | 400 | 139 | 74 | 54 | 267 | 133 | 25 | 0 | 1 | 0 | 1 |
| **2000** | 165 | 122 | 107 | 394 | 152 | 70 | 67 | 289 | 105 | 23 | 0 | 0 | 1 | 1 |
| **2001** | 230 | 140 | 101 | 471 | 214 | 82 | 72 | 368 | 103 | 24 | 0 | 0 | 1 | 1 |
| **2002** | 179 | 116 | 87 | 382 | 166 | 69 | 61 | 296 | 86 | 18 | 0 | 1 | 1 | 2 |
| **2003** | 156 | 106 | 93 | 355 | 149 | 77 | 70 | 296 | 59 | 14 | 0 | 0 | 0 | 0 |
| **2004** | 194 | 117 | 121 | 432 | 192 | 78 | 87 | 357 | 75 | 22 | 0 | 0 | 0 | 0 |
| **2005** | 228 | 125 | 121 | 474 | 221 | 88 | 82 | 391 | 83 | 20 | 0 | 0 | 0 | 0 |
| **2006** | 224 | 114 | 120 | 458 | 218 | 69 | 86 | 373 | 85 | 28 | 0 | 0 | 0 | 0 |
| **2007** | 212 | 123 | 101 | 436 | 207 | 86 | 69 | 362 | 74 | 36 | 0 | 0 | 0 | 0 |
| **2008** | 263 | 111 | 113 | 487 | 255 | 72 | 79 | 406 | 81 | 31 | 0 | 0 | 0 | 0 |
| **2009** | 214 | 129 | 130 | 473 | 201 | 85 | 96 | 382 | 91 | 18 | 4 | 1 | 3 | 8 |
| **2010** | 177 | 85 | 97 | 359 | 175 | 61 | 64 | 300 | 59 | 23 | 0 | 1 | 1 | 2 |
| **2011** | 212 | 93 | 141 | 446 | 174 | 57 | 76 | 307 | 139 | 16 | 13 | 8 | 12 | 33 |
| **2012** | 250 | 122 | 143 | 515 | 130 | 47 | 51 | 228 | 287 | 14 | 54 | 14 | 29 | 97 |
| **2013** | 234 | 135 | 156 | 525 | 144 | 70 | 80 | 294 | 231 | 17 | 44 | 18 | 21 | 83 |
| **2014** | 218 | 123 | 133 | 474 | 205 | 69 | 80 | 354 | 120 | 9 | 3 | 3 | 9 | 15 |
| **2015** | 218 | 80 | 135 | 433 | 207 | 55 | 96 | 358 | 75 | 10 | 2 | 0 | 5 | 7 |
| **2016** | 178 | 96 | 144 | 418 | 173 | 67 | 84 | 324 | 94 | 11 | 0 | 0 | 1 | 1 |
| **2017** | 215 | 92 | 135 | 442 | 202 | 50 | 96 | 348 | 94 | 10 | 0 | 3 | 1 | 4 |
| **2018** | 242 | 106 | 157 | 505 | 236 | 75 | 94 | 405 | 100 | 6 | 0 | 0 | 2 | 2 |
| **2019** | 243 | 88 | 151 | 482 | 237 | 63 | 99 | 399 | 83 | 7 | 1 | 5 | 5 | 11 |
| **Sum** | **5800** | **3373** | **3247** | **12420** | **4705** | **1633** | **1928** | **8266** | **4154** | **521** | **121** | **58** | **92** | **271** |

Table 9: Shown are the number of articles found in PRL per year (in blue section) and subdivided into the PRL subject areas considered here. The "Total" is simply their sum. The green shaded area shows the results for articles on the arXiv with a found reference to a PRL publication. The red column is the number of PRL articles without a found reference on the arXiv. The grey columns are arXiv articles with a reference to PRL, but to a subject area that is not *GA, NP* or *EPF*. The yellow shaded area shows the number of articles found on the arXiv, corresponding to a PRL publication and found through title matching as described above. More in text below.

It is clear from Table 9, by simple looking at the red column "#not on arX" that we see the same ramping up of journal references on the arXiv in the early years of the arXiv. From then on, most PRL articles exist on arXiv with a journal reference. There is a dip for about 4 years from 2011 to 2014. More about this in the figures below.

Before turning to more results from Table 9 a little bit more detail about the title matching. The title match algorithm is applied to all PRL articles that do not have a reference on the arXiv (4154 publications). The PRL title is compared to all arXiv titles in the *arXiv-without-journal* dataset. This dataset contains a total of 249174 arXiv titles from all repositories considered. For each PRL title a match is performed, with the algorithm outlined above, and a total of 271 matches are found. This corresponds to 5.7 $10^7$ attempted title matches.

Before looking at these matched titles it is important to check how many random title matches there are. To do this, the same PRL data set is used (ones without an arXiv reference, 4154), but now these titles are matched to the *arXiv-with-journal* dataset. This dataset, with 254472 titles, also includes all arXiv titles with a reference to PRL. A total of 5.5 $10^7$ title matches are performed and 13 matches are found. In an ideal world one expects zero. Of the 13 matches 8 were some form of a reference to PRL. However, the reference is incomplete. These are PRL references on the arXiv, that the algorithm decoding the journal reference on the arXiv, could not convert into a correct reference to an existing PRL publication. Five of the 13 title matches found were references to other journals. This number is very small compared to the number of matches tried, so the conclusion is that the sample of 271 matches above is pure and these are indeed PRL publications.

Below are the results of Table 9 displayed graphically, in the same manner as was done for PLB above. For a more detailed explanation of the figures see that section.

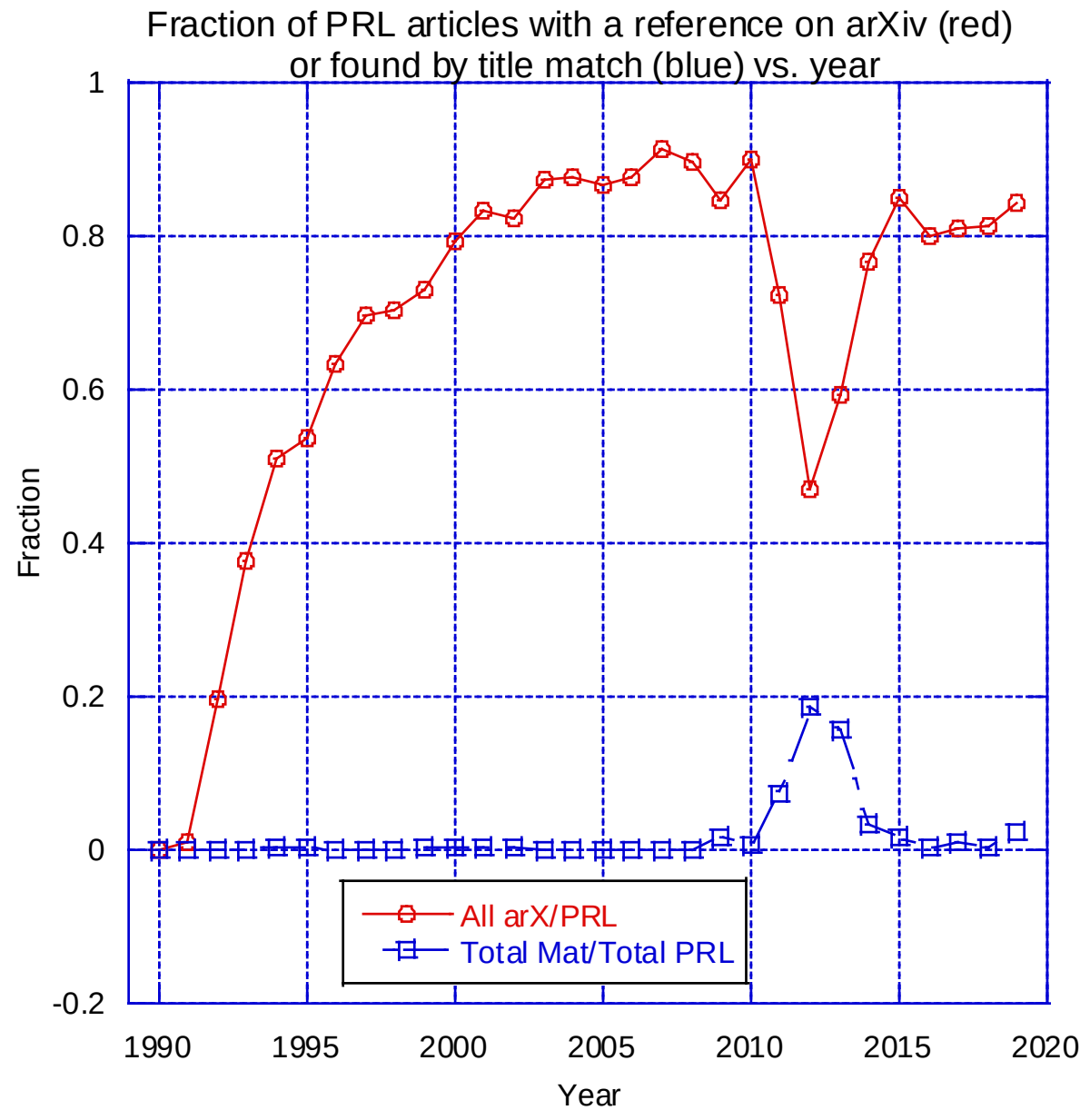

Figure 7: Fraction of PRL articles with journal reference, found on arXiv (red) or by title match (blue)

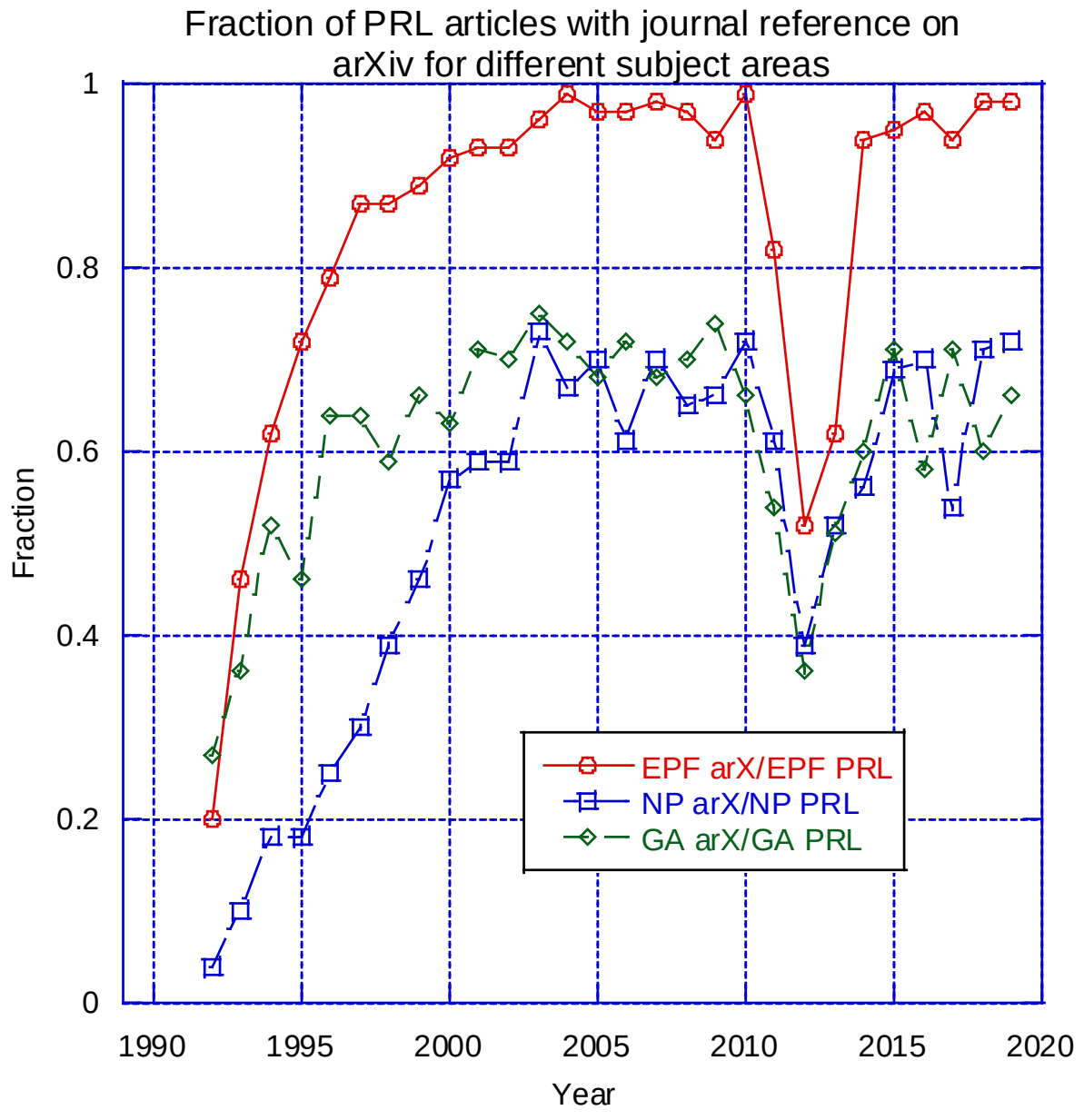

Figure 8: Same as Figure 3, but now for the different PRL subject areas

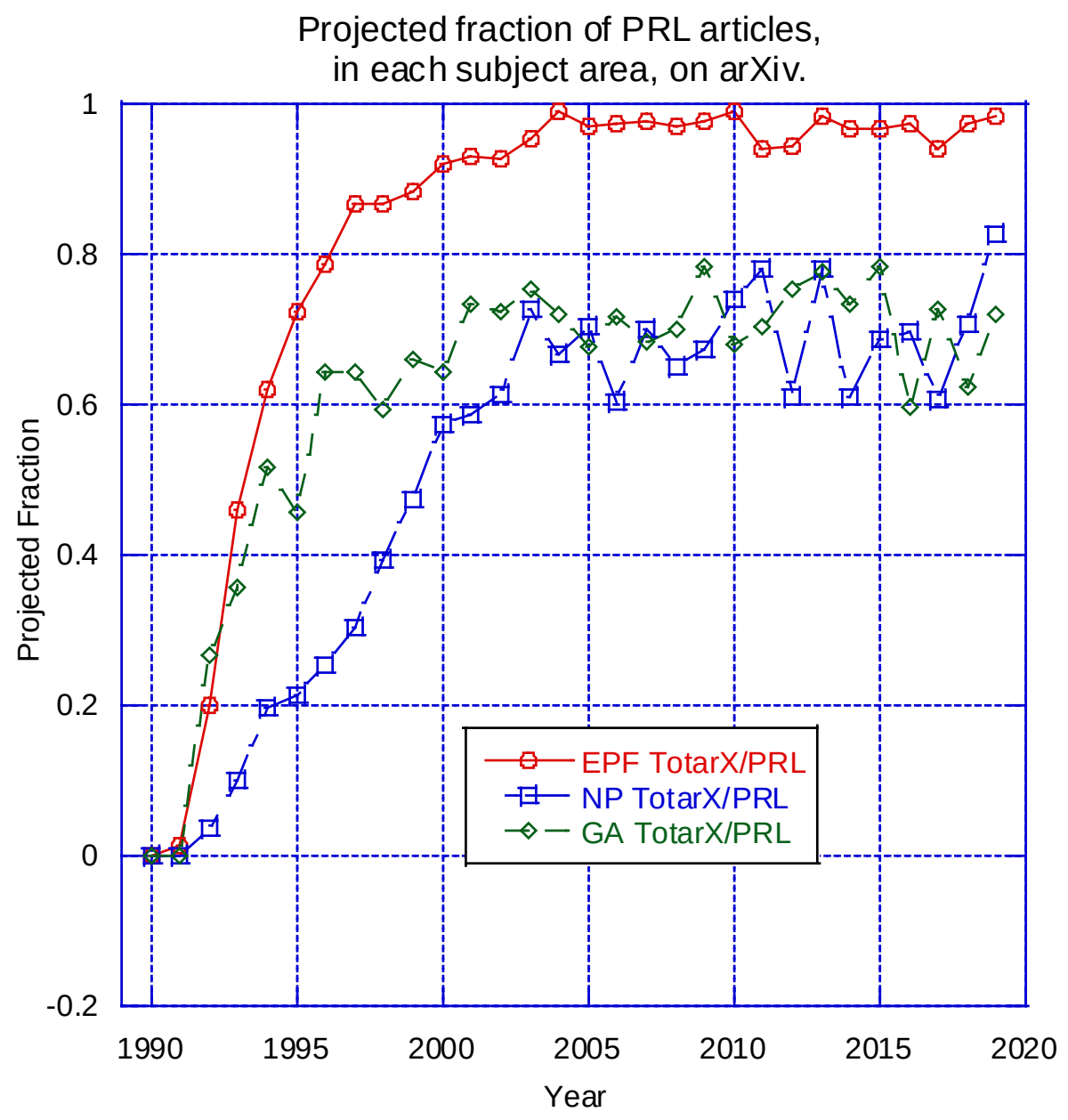

Figure 9: Projected fraction of PRL articles with corresponding arXiv article for subject areas ( see text for more detail)

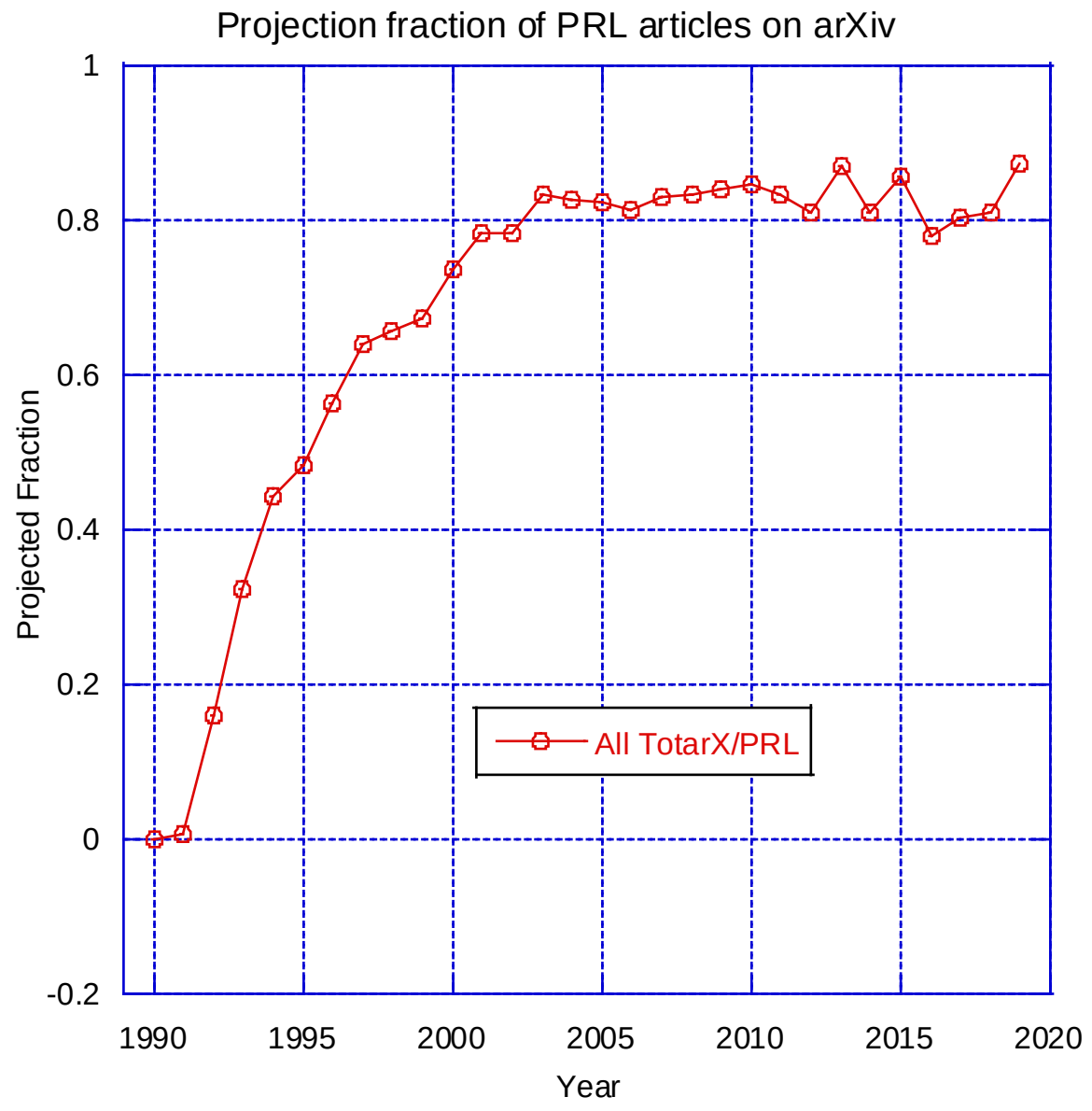

Figure 10: Same as Figure 5, but now for all PRL articles.

*Observations and conclusions*

The following conclusions can be drawn from these numbers and graphs:

1. The different arXiv repositories start around 1990-1991. It took about 10 years before the community, publishing in PRL, completely used the arXiv as its preprint server (Figure 7).
2. When trying to associate articles in PRL with their corresponding “preprint” version on the arXiv, the journal reference exists in most cases for all years, except for a four-year dip from 2011 to 2015 and then recovers (red curve in Figure 7). This is very different from PLB, as seen in Figure 3.
3. Looking at the subject areas in PRL, basically the same conclusion can be drawn (Fig 8). The major difference is that in general the subject area EPF has a larger fraction of journal references on the arXiv.
4. Figures 7 also shows the title match articles found (blue curve). The articles found are all in the area where journal references are missing.
5. Taking into account the efficiency of the title matching, one can calculate the total number of PRL articles with a corresponding article on the arXiv. Figure 9 shows the projected fraction of PRL publications with a corresponding article on the arXiv for the subject areas. The fraction for all articles is shown in Fig. 10.
6. The conclusion is that the missing journal references in the dip (2011-2015), correspond to PRL publications with a corresponding arXiv article, but it has no journal reference.
7. It seems that the subject areas GA and NP reach a plateau, in terms of having corresponding articles on the arXiv. We think this is correct and is probably due to the way the arXiv articles were selected from particular arXiv repositories. Most likely additional repositories will need to be included.

**Overall observations and conclusions.**

This report establishes a correlation between articles published in the two journals Physics Letters B (PLB) and Physical Review Letters and those articles on the arXiv, in the fields of Nuclear Physics, Particle Physics and Astrophysics. Here the assumption is that the arXiv contains the "preprint" of each of those journal articles. The journal articles themselves have no reference to such "preprints", however the arXiv in principle maintains a journal reference in each of its articles, if they are published in a journal. This information was used to try and establish a link between each published journal article and an arXiv article, using the arXiv repositories for *hep-ex, hep-ph, hep-th, hep-lat, astro-ph, nucl-ex* and *nucl-th.*

It was found that after the arXiv started around 1990/91 it took about 10 years before nearly all journal articles in PLB and PRL exist in preprint form on the arXiv and the arXiv articles have a corresponding journal reference. This continues to about 2010. Starting in 2011, for both journals there is a ~50% drop in journal references. For PRL this is a dip and the journal references on the arXiv recover to previous levels in 2016. For PLB the drop starts in 2011, but there is no recovery and it continues to fall to about 30% in 2019. This behavior is seen in all the journal subject areas considered for both journals.

The question then is: Do the journal articles, with no journal reference on the arXiv, have a corresponding article on the arXiv or is it simply not existing on the arXiv? To address this the journal articles, without a corresponding journal reference in an arXiv article, were attempted to be matched to arXiv articles (without journal references). This was done by a title matching algorithm developed for that purpose. Matches were found in the time periods where journal references were missing. The conclusion is that all PLB journal articles have a corresponding arXiv article, but in many cases without a journal reference. For PRL the same can be concluded for the subject area *Elementary Particles and Fields*. For *Gravitation and Astrophysics* and *Nuclear Physics* about 80% of the journal articles have a corresponding article in the arXiv repositories considered. A similar result (but less pronounced) is observed for the subject areas *Experiments* and *Astrophysics and Cosmology* in PLB. Although not proven, we speculate that the missing arXiv articles are in arXiv repositories not considered here.

*Acknowledgements & disclaimer:*
The support from COFI is acknowledged, allowing me to use the facility in San Juan, PR for a while, enabling the download of webpages and initial development of the scripts used. This work was not supported by external funding and simply has arisen from a personal interest.
The information in this report may be available via other means and there is no claim of it being unique. This information may exist in some other form, but I was not able to easily find it. I simply pursued an interest in getting a historical perspective of publications in the field of Particle Physics. Data used for this report are available to anybody upon request.

**Appendix A:** The acronyms and name of journals.

To interpret and keep track of the journal reference of an article on the arXiv, each journal reference is reduced to an abbreviation (called *Journal-ID*), consisting of the first four capital letters found in the reference. These journal abbreviations are then assigned to a new short name (called *Journal-group*), where it is possible to assign multiple *Journal_ID*'s to one *Journal-group*. This allows easy assignment of different representations of journal references to be assigned to one *Journal-group*. The text below is simply the output of collecting all journal references in the hep-ex arXiv repository. When a *Journal-ID* is found for the first time, it is listed as well as the actual first reference. For more explanation see below.

For more explanations and examples, see further below.

```
Recognized CONF  used as  Conf | Nucl.Phys.Proc.Suppl.35:261-263,1994
Recognized NPB  used as  NPB | Nucl.Phys.B439:471-502,1995
Recognized NCA  used as  NuovoCim | NuovoCim.A108:1035-1040,1995
Recognized PPNP  used as  Review | Prog.Part.Nucl.Phys.36:437-446,1996
Recognized APPB  used as  Acta | Acta Phys.Polon.B28:1155-1158,1997
Recognized NIMB  used as  NIM | Nucl.Instrum.Meth.B119:253-258,1996
Recognized JPG  used as  JPhyG | J.Phys.G22:797-814,1996
Recognized ARNP  used as  Review | Ann.Rev.Nucl.Part.Sci.46:533-608,1996
Recognized RMP  used as  Review | Rev.Mod.Phys.69:137-212,1997
Recognized IJMP  used as  IJMP | Int.J.Mod.Phys.A12:3827-3836,1997
Recognized PRC  used as  PRC | Phys.Rev.C56:2774-2778,1997
Recognized EPJC  used as  EPJC | Eur.Phys.J.C1:509-513,1998
Recognized PAN  used as  PAN | Phys.Atom.Nucl.61:66-73,1998; Yad.Fiz.61:72-79,1998
Recognized NPA  used as  NPA | Nucl.Phys. A638 (1998) 249c-260c
Recognized LNP  used as  Review | Lect.Notes Phys.499:43-56,1997
Recognized EJPA  used as  EPJA | Eur.J.Phys.A1:299-306,1998
Recognized AP  used as  Astro | Astropart.Phys.10:11-20,1999
Recognized CJP  used as  Acta | Czech.J.Phys. 49S2 (1999) 119-126
Recognized EPJA  used as  EPJA | Eur.Phys.J.A5:441-443,1999
Recognized JHEP  used as  JHEP | JHEP 9908:004,1999
Recognized NJP  used as  New J Phys | NewJ.Phys.2:1,2000
Recognized NCC  used as  NuovoCim | Nuovo Cim.C24:761-770,2001
Recognized ASS  used as  Astro | Astrophys.Space Sci.282:235-244,2002
Recognized P  used as  Acta | Pramana 62:561-564,2004
Recognized APS  used as  Acta | ActaPhys.Slov.55:15-24,2005
Recognized APHA  used as  Acta | Acta Phys.Hung. A24 (2005) 321-328
Recognized RPP  used as  Review | Rept.Prog.Phys. 68 (2005) 2773-2828
Recognized JINS  used as  JINST | JINST 1:P07002,2006
Recognized APPS  used as  Acta | Acta Phys.Polon.Supp.1:257-260,2008
Recognized NCB  used as  NuovoCim | Nuovo Cim.B123:409-414,2008
Recognized NIMP  used as  NIM | Nuclear Instruments and Methods in Physics Research
Section A:   Accelerators, Spectrometers, Detectors and Associated Equipment, Volume
654,   Issue 1, 21 October 2011, Pages 481-489
Recognized PLBV  used as  PLB | Physics Letters B, Volume 698, Issue 3, 11 April 2011,
Pages   196-218
Recognized PRLV  used as  PRL | Physical Review Letters, Vol.107, No.21, 2011
Recognized NIM  used as  NIM | Nucl.Inst.Meth. 676 (2012) 44-49
Recognized TEPJ  used as  EPJC | The European Physical Journal C - Particles and
Fields, 2012,   Volume 72, Number 4, 1973
Recognized INCC  used as  NuovoCim | Il Nuovo Cimento C 36 01 (2013)
Recognized SCPM  used as  China | Sci. China-Phys. Mech. Astron. 60, 071011 (2017)
Recognized CJPV  used as  China | Chinese Journal of Physics, Volume 58 (2019) 63-74
```

Example: Using the last line in the above text, the *Journal-ID* “SCPM”, corresponding to the journal listed, is assigned to the *Journal-group* ‘China’. This group was basically used to identify journals published in China. The *Journal-ID* “CJPV” is assigned to the same *Journal-group*.
The *Journal-ID* ‘CONF’ was special. The arXiv contains many journal reference to proceedings, symposia, special events etc. This *Journal-ID* was assigned to a journal reference if the reference contained one of the following strings: 'conf', 'CONF', 'Conf', 'Proc', 'PoS', 'Pos', 'PROC', 'work', 'Works', 'Renc', 'Serie', 'IEEE', 'ICFA', 'proc' or 'Symp'.

**Appendix B**: The contents of the arXiv repository *nucl-th*, showing in which journals articles are published.
This table is similar to the one displayed in the text above for the repository *hep-ex*. The obvious difference is that the field of nuclear physics theory publishes in different journals than experimental particle physics (*hep-ex*).

| year | PTP | PRD | FBS | PTEP | JHEP | EPJC | EPJA | HIP | PLB | PRL | PRA | PRC | NPA | China | JPhyG | NIM | Astro | IJMP | Conf | Review | Acta | PAN | Other |
|---|---|---|---|---|---|---|---|---|---|---|---|---|---|---|---|---|---|---|---|---|---|---|---|
| 1992 | 0 | 0 | 0 | 0 | 0 | 0 | 0 | 0 | 1 | 4 | 0 | 5 | 0 | 0 | 1 | 0 | 0 | 2 | 0 | 0 | 0 | 0 | 3 |
| 1993 | 1 | 4 | 1 | 0 | 0 | 0 | 0 | 0 | 23 | 8 | 0 | 62 | 8 | 0 | 0 | 0 | 1 | 2 | 2 | 1 | 0 | 0 | 3 |
| 1994 | 6 | 11 | 4 | 0 | 7 | 0 | 0 | 0 | 41 | 19 | 0 | 91 | 48 | 1 | 6 | 1 | 4 | 3 | 0 | 1 | 1 | 1 | 10 |
| 1995 | 2 | 9 | 13 | 0 | 11 | 0 | 0 | 1 | 53 | 20 | 7 | 132 | 47 | 0 | 5 | 0 | 4 | 1 | 2 | 4 | 2 | 6 | 15 |
| 1996 | 5 | 11 | 3 | 0 | 11 | 0 | 0 | 5 | 69 | 26 | 0 | 162 | 98 | 1 | 7 | 0 | 3 | 4 | 2 | 5 | 10 | 5 | 29 |
| 1997 | 8 | 6 | 3 | 0 | 27 | 0 | 0 | 4 | 68 | 25 | 4 | 215 | 107 | 8 | 16 | 0 | 3 | 5 | 2 | 9 | 3 | 11 | 42 |
| 1998 | 10 | 11 | 9 | 0 | 0 | 8 | 16 | 2 | 97 | 32 | 0 | 241 | 164 | 1 | 15 | 4 | 6 | 9 | 9 | 7 | 23 | 6 | 29 |
| 1999 | 10 | 11 | 31 | 0 | 0 | 5 | 30 | 5 | 87 | 42 | 2 | 275 | 114 | 3 | 27 | 0 | 4 | 6 | 13 | 16 | 12 | 3 | 37 |
| 2000 | 11 | 13 | 19 | 0 | 0 | 4 | 43 | 1 | 79 | 35 | 1 | 273 | 129 | 1 | 17 | 2 | 2 | 7 | 11 | 9 | 11 | 6 | 30 |
| 2001 | 10 | 9 | 9 | 0 | 0 | 4 | 25 | 7 | 60 | 38 | 2 | 262 | 171 | 2 | 40 | 1 | 5 | 14 | 19 | 11 | 21 | 12 | 39 |
| 2002 | 8 | 8 | 5 | 0 | 1 | 1 | 33 | 5 | 58 | 32 | 2 | 349 | 131 | 2 | 30 | 0 | 7 | 4 | 12 | 6 | 17 | 9 | 42 |
| 2003 | 23 | 11 | 26 | 0 | 0 | 10 | 35 | 14 | 51 | 39 | 0 | 326 | 167 | 8 | 27 | 0 | 4 | 19 | 28 | 16 | 11 | 11 | 42 |
| 2004 | 30 | 10 | 9 | 0 | 0 | 4 | 57 | 7 | 73 | 29 | 2 | 343 | 111 | 8 | 59 | 3 | 3 | 22 | 11 | 30 | 17 | 10 | 56 |
| 2005 | 18 | 4 | 4 | 0 | 0 | 6 | 76 | 0 | 52 | 42 | 4 | 346 | 118 | 7 | 50 | 1 | 9 | 23 | 27 | 13 | 22 | 6 | 51 |
| 2006 | 12 | 3 | 5 | 0 | 0 | 5 | 32 | 0 | 64 | 28 | 2 | 347 | 93 | 7 | 34 | 0 | 3 | 22 | 51 | 19 | 52 | 7 | 51 |
| 2007 | 25 | 10 | 5 | 0 | 0 | 11 | 57 | 0 | 69 | 37 | 1 | 342 | 119 | 5 | 52 | 3 | 8 | 36 | 78 | 18 | 10 | 7 | 70 |
| 2008 | 14 | 8 | 28 | 0 | 0 | 3 | 42 | 0 | 55 | 34 | 3 | 335 | 75 | 7 | 63 | 0 | 6 | 24 | 53 | 21 | 16 | 10 | 58 |
| 2009 | 13 | 8 | 6 | 0 | 0 | 9 | 32 | 0 | 48 | 43 | 1 | 389 | 86 | 20 | 46 | 1 | 3 | 32 | 62 | 21 | 31 | 7 | 53 |
| 2010 | 24 | 6 | 8 | 0 | 0 | 2 | 35 | 0 | 39 | 32 | 0 | 419 | 53 | 7 | 73 | 0 | 2 | 32 | 69 | 14 | 13 | 1 | 47 |
| 2011 | 16 | 15 | 25 | 0 | 1 | 3 | 22 | 0 | 52 | 32 | 0 | 345 | 42 | 6 | 15 | 1 | 8 | 29 | 100 | 7 | 7 | 6 | 50 |
| 2012 | 17 | 11 | 2 | 3 | 1 | 1 | 16 | 0 | 24 | 20 | 1 | 219 | 23 | 7 | 16 | 0 | 2 | 19 | 67 | 12 | 11 | 4 | 44 |
| 2013 | 0 | 21 | 12 | 3 | 0 | 3 | 19 | 0 | 44 | 32 | 2 | 250 | 35 | 9 | 20 | 0 | 6 | 23 | 43 | 6 | 2 | 3 | 57 |
| 2014 | 0 | 25 | 6 | 8 | 0 | 6 | 22 | 0 | 33 | 33 | 1 | 319 | 28 | 5 | 19 | 0 | 5 | 16 | 43 | 8 | 10 | 2 | 51 |
| 2015 | 0 | 35 | 0 | 9 | 0 | 1 | 16 | 0 | 24 | 24 | 0 | 301 | 22 | 4 | 30 | 0 | 1 | 16 | 34 | 5 | 3 | 2 | 66 |
| 2016 | 0 | 47 | 4 | 6 | 0 | 5 | 45 | 0 | 16 | 26 | 1 | 341 | 18 | 5 | 17 | 0 | 2 | 19 | 47 | 9 | 4 | 3 | 77 |
| 2017 | 0 | 46 | 11 | 9 | 0 | 9 | 26 | 0 | 29 | 23 | 2 | 396 | 18 | 12 | 16 | 0 | 0 | 20 | 36 | 5 | 11 | 0 | 84 |
| 2018 | 0 | 51 | 5 | 3 | 1 | 4 | 12 | 0 | 34 | 26 | 0 | 440 | 10 | 7 | 11 | 0 | 1 | 13 | 25 | 11 | 5 | 3 | 76 |
| 2019 | 0 | 49 | 4 | 6 | 1 | 9 | 19 | 0 | 16 | 21 | 0 | 354 | 18 | 8 | 19 | 1 | 3 | 11 | 32 | 6 | 6 | 0 | 51 |
| 2020 | 0 | 1 | 0 | 1 | 0 | 0 | 1 | 0 | 3 | 2 | 0 | 12 | 2 | 2 | 0 | 0 | 0 | 0 | 0 | 1 | 0 | 0 | 5 |
| 9999 | 0 | 0 | 0 | 0 | 0 | 2 | 0 | 0 | 1 | 0 | 0 | 2 | 0 | 0 | 1 | 0 | 0 | 0 | 6 | 1 | 0 | 0 | 34 |

Table B1: The contents of the *nucl-th* repository, after selecting all articles from the *arXiv-with-journal* set ( after Step 3), interpreting that journal reference and identifying the journal and the year of publication in the journal. Some more explanations are in the text. The abbreviations for the journal names are explained in Appendix A.

**Appendix C:** Contents and short history of PRL

Physical review Letters (PRL) has been published since 1958 [3] and all its contents are available on WEB pages. A summary of the number of articles per year is in Table C1 below. In 1970 PRL introduced subject areas, to associate published articles with research areas of physics, since PRL publishes all areas of physics, unlike nearly all other letter journals. For this report we use only the subject areas: *Gravitation and Astrophysics*(GA), *Elementary Particles & Fields*(EPF) and *Nuclear Physics*(NP). As Table C1 shows Nuclear Physics was initially called "Nuclei", but that name was phased out over a few years from 1985 through 1987 and replaced by Nuclear Physics. For this report we are only considering the years after 1990, so by that time the name "Nuclei" was no longer in use.

All the contents of PRL webpages were downloaded at the end of 2019 and all numbers presented here reflect the status at that time. Changes to the webpages after that time are not included in this report. The PRL webpages have special features for "*Editors Suggestions*", "*Featured in Physics*" under the heading "*Highlighted Articles*". That section is simply skipped and only the contents of the section called "*Letters*" is used in this report. Sections called "*Errata*" or "*Comments*" or any other names over the years are all skipped.

Typically, a Volume in PRL covers about 6 months of the year and has on the order of 25-27 issues. Each web page covers one volume. The total time span of 1958 through 2019 is covered by volumes 1 through 123.

Articles were identified by Volume and page number up to and including Volume 86 in 2001. Starting with Volume 87, the article ID was introduced and page numbers were no longer used.

| Year | EPF PRL | NP PRL | Nuclei PRL | GA PRL | Other PRL | None PRL | Total PRL |
|---|---|---|---|---|---|---|---|
| **1958** | | | | | | 206 | 206 |
| **1959** | | | | | | 413 | 413 |
| **1960** | | | | | | 389 | 389 |
| **1961** | | | | | | 409 | 409 |
| **1962** | | | | | | 398 | 398 |
| **1963** | | | | | | 414 | 414 |
| **1964** | | | | | | 591 | 591 |
| **1965** | | | | | | 765 | 765 |
| **1966** | | | | | | 877 | 877 |
| **1967** | | | | | | 948 | 948 |
| **1968** | | | | | | 1120 | 1120 |
| **1969** | | | | | | 974 | 974 |
| **1970** | 125 | 0 | 81 | | 339 | 470 | 1015 |
| **1971** | 255 | 0 | 157 | | 695 | | 1107 |
| **1972** | 201 | 0 | 155 | | 746 | | 1102 |
| **1973** | 178 | 0 | 124 | | 598 | | 900 |
| **1974** | 204 | 0 | 120 | | 617 | | 941 |
| **1975** | 256 | 0 | 128 | | 697 | | 1081 |
| **1976** | 190 | 0 | 104 | | 737 | | 1031 |
| **1977** | 157 | 0 | 146 | | 648 | | 951 |
| **1978** | 168 | 0 | 214 | | 652 | | 1034 |
| **1979** | 162 | 0 | 171 | | 744 | | 1077 |

| Year | EPF PRL | NP PRL | Nuclei PRL | GA PRL | Other PRL | None PRL | Total PRL |
|---|---|---|---|---|---|---|---|
| **1979** | 162 | 0 | 171 | | 744 | | 1077 |
| **1980** | 181 | 0 | 138 | | 772 | | 1091 |
| **1981** | 111 | 0 | 140 | | 779 | | 1030 |
| **1982** | 124 | 0 | 115 | | 874 | | 1113 |
| **1983** | 164 | 0 | 111 | | 978 | | 1253 |
| **1984** | 161 | 0 | 136 | | 1058 | | 1355 |
| **1985** | 226 | 18 | 122 | | 1238 | | 1604 |
| **1986** | 223 | 21 | 85 | | 1394 | | 1723 |
| **1987** | 177 | 93 | 12 | | 1351 | | 1633 |
| **1988** | 157 | 94 | 0 | 62 | 1328 | | 1641 |
| **1989** | 167 | 102 | 0 | 66 | 1344 | | 1679 |
| **1990** | 171 | 85 | 0 | 62 | 1506 | | 1824 |
| **1991** | 133 | 100 | 0 | 61 | 1637 | | 1931 |
| **1992** | 159 | 106 | 0 | 64 | 1769 | | 2098 |
| **1993** | 163 | 110 | 0 | 73 | 1918 | | 2264 |
| **1994** | 124 | 106 | 0 | 58 | 1801 | | 2089 |
| **1995** | 155 | 131 | 0 | 81 | 2284 | | 2651 |
| **1996** | 140 | 122 | 0 | 87 | 2370 | | 2719 |
| **1997** | 165 | 112 | 0 | 87 | 2338 | | 2702 |
| **1998** | 181 | 117 | 0 | 106 | 2680 | | 3084 |
| **1999** | 157 | 161 | 0 | 82 | 2469 | | 2869 |
| **2000** | 165 | 122 | 0 | 107 | 2678 | | 3072 |

| Year | EPF PRL | NP PRL | Nuclei PRL | GA PRL | Other PRL | None PRL | Total PRL |
|---|---|---|---|---|---|---|---|
| **2001** | 230 | 140 | 0 | 101 | 2775 | | 3246 |
| **2002** | 179 | 116 | 0 | 87 | 2603 | | 2985 |
| **2003** | 156 | 106 | 0 | 93 | 2795 | | 3150 |
| **2004** | 194 | 117 | 0 | 121 | 3326 | | 3758 |
| **2005** | 228 | 125 | 0 | 121 | 3364 | | 3838 |
| **2006** | 224 | 114 | 0 | 120 | 3461 | | 3919 |
| **2007** | 212 | 123 | 0 | 101 | 3253 | | 3689 |
| **2008** | 263 | 111 | 0 | 113 | 3551 | | 4038 |
| **2009** | 214 | 129 | 0 | 130 | 3078 | | 3551 |
| **2010** | 177 | 85 | 0 | 97 | 2890 | | 3249 |
| **2011** | 212 | 93 | 0 | 141 | 2899 | | 3345 |
| **2012** | 250 | 122 | 0 | 143 | 3356 | | 3871 |
| **2013** | 234 | 135 | 0 | 156 | 3132 | | 3657 |
| **2014** | 218 | 123 | 0 | 133 | 2375 | | 2849 |
| **2015** | 218 | 80 | 0 | 135 | 2111 | | 2544 |
| **2016** | 178 | 96 | 0 | 144 | 1972 | | 2390 |
| **2017** | 215 | 92 | 0 | 135 | 2063 | | 2505 |
| **2018** | 242 | 106 | 0 | 157 | 2301 | | 2806 |
| **2019** | 243 | 88 | 0 | 151 | 2198 | | 2680 |
| **Sums** | **9549** | **3701** | **2430** | **3375** | **95286** | **7974** | **122315** |

Table C1: The number of publication in PRL for each year, as found on the PRL web pages. Only the subject areas *Gravitation & Astrophysics* (GA), *Elementary Particles & Fields* (EPF) and *Nuclear Physics* (NP) are listed. All other subject areas are grouped into the class "Other". These subject areas were introduced in 1970 and have been used ever since.